\documentclass[dvips]{aa}     
\usepackage{graphicx}

\font\tenbi=cmmib10
\newfam\bifam  \textfont\bifam=\tenbi

\def\etal{\mbox{et al.~}}

\begin{document}
 
\thesaurus{06(03.13.2; 03.13.6; 03.13.5; 06.15.1)}

\title{The art of fitting p-mode spectra: Part II. Leakage and noise 
covariance matrices}
\author{Thierry~Appourchaux\inst{1}, Maria-Cristina 
Rabello-Soares\inst{1}, Laurent~Gizon\inst{1,2}}
\institute{Space Science Department of ESA, ESTEC, NL-2200 AG 
Noordwijk
\and
W.W.Hansen Experimental Physics Laboratory, Center for Space Science 
and 
Astrophysics, Stanford University, Stanford, CA 94305-4085, USA}
\offprints{thierrya\mbox{@}so.estec.esa.nl}

\date{Received / Accepted}

\authorrunning{Appourchaux \etal}
\titlerunning{The art of fitting, Part II}

\maketitle
\begin{abstract}
In Part I we have developed a theory 
for fitting p-mode Fourier spectra assuming that these spectra have a 
multi-normal distribution.  We showed, using Monte-Carlo simulations, 
how one can obtain p-mode parameters using 'Maximum Likelihood 
Estimators'.  In this article, hereafter Part II, we show how to use 
the theory 
developed in Part I for fitting real data.  We introduce 4 new 
diagnostics in helioseismology: the $(m,\nu)$ echelle 
diagramme, the cross echelle diagramme, the inter echelle diagramme, 
and the ratio cross spectrum.
These diagnostics are extremely powerful to visualize and understand 
the covariance matrices of the Fourier spectra, and also to find 
bugs in the data analysis code.  These diagrammes can also be used to 
derive 
quantitative information on the mode leakage and noise covariance 
matrices.  Numerous examples using 
the LOI/SOHO and GONG data are given.
\end{abstract}

\section{Introduction}
The physics of the solar interior is known from inversion of solar 
p-mode frequencies and splittings.  These measurements are derived 
from fitting p-mode
Fourier spectra. Schou (1992) was the first one to assume a 
multi-normal distribution for p-mode Fourier spectra and using a real 
leakage matrix.  Following this pioneering 
work, Appourchaux \etal (1997) (hereafter Part I), generalized 
the the\-o\-re\-ti\-cal back\-ground for fitting 
p-mode Fou\-rier spectra to complex leakage matrix, and included 
explicitly the correlation of the noise between the Fourier spectra.  
Using Monte-Carlo simulations, we showed 
that 
our fitted parameters were unbiased.  We also studied systematic 
errors due to an imperfect knowledge of the leakage  covariance 
matrix.    
Unfortunately, a theoretical knowledge of 
fitting data is not enough as only real data will teach us if our 
approach is correct.  Contrary to fitting p-mode power spectra, the 
process of fitting the Fourier spectra as described in 
Part I is rather difficult to understand and visualize.  Schou (1992) 
gave 
a few diagnostics for understanding how the Fourier spectra are 
fitted but without 
showing an easy way to visualize the covariance matrices.

In this paper, we show how one can easily visualize 
the mode and noise correlation matrices, and then derive the mode 
leakage matrix.  In the first 
section, we describe 4 new diagrammes that have various diagnostics 
power.  In the 
second section, we describe how we use those diagrammes for inferring
 the leakage and noise covariance matrices for the data of the 
Luminosity Oscillations Imager (LOI) on board the Solar and 
Heliospheric Observatory (SOHO) 
data (Appourchaux et al, 1997), and for the data of the Global 
Oscillations Network Group (GONG) (Hill et al, 1996).  
The LOI time series starts on 27 March 1996 and ends on 27 March 
1997 with a duty cycle greater than 99\%.  The 
GONG time series starts on 27 August 1995 and ends on 22 August 
1996 with a 75\% duty cycle. In the 
last 
section we conclude by emphasizing the usefulness of these 
diagrammes. 

\begin{figure}[t]
\centering
\includegraphics[angle=90,width=7.5cm]{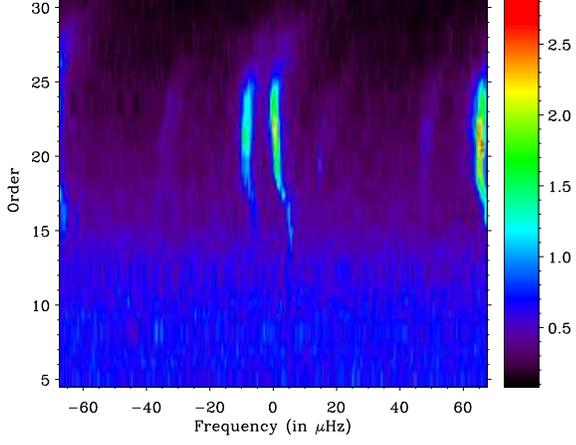}
\caption{Amplitude spectra echelle diagramme for 1 year of LOI data 
seeing the Sun as a 
star.  The scale is in part-per-million\,$\mu$Hz$^{-1/2}$ 
(ppm\,$\mu$Hz$^{-1/2}$).  
The spacing is tuned for $l=0$.  The $l=0$ modes are at the center, 
the $l=2$ modes are about 
10 $\mu$Hz on the left hand side, the $l=1$ modes are about +65 
$\mu$Hz on the right hand side, the $l=3$ can be faintly seen at 
about 
12 $\mu$Hz from the left hand side of the $l=1$.  Other modes such as 
$l=4$ and $l=5$ can also be
seen faintly seen at -35 $\mu$Hz and +15 $\mu$Hz, respectively.  The 
distortion of the ridges are due to 
sound speed gradients in the solar core.}
\label{echelle} 
\end{figure}

\section{Diagnostics for helioseismic data analysis}
The echelle diagramme was first introduced by Grec (1981).  
It is based on the fact that the low-degree modes are essentially 
equidistant in frequency for a given $l$; the typical spacing for 
$l=0$ is 136 $\mu$Hz.  The spectrum is cut into 
pieces of 136 $\mu$Hz which are stacked on top of each other.  
Since the modes are not truly equidistant in frequency, the echelle 
diagramme 
shows up power as distorted ridges; an example is given in Fig. 
\ref{echelle} for the LOI/SOHO instrument seeing the Sun as a star.

Another useful diagramme was introduced by Brown (1985), the 
so-called $(m,\nu)$ diagramme which shows how the frequency of an 
$l,m$ mode depends upon $m$.  Most often this diagramme is only shown 
for 
a single $n$ and for intermediate degrees $l \ge 10$.

The purpose of these diagrammes is always to show an estimate of the 
variance of the spectra.
In our case we also want to visualize not only the variance but also 
the covariance of the 
Fourier spectra.  Here we briefly recall from Part I that the 
observed Fourier spectra ($\vec{y}$) can be related to the
individual Fourier spectra of the normal modes ($\vec{x}$) by the 
leakage 
matrix $\tens{\cal{C}}^{(l,l')}$ by:
\begin{equation}
\vec{y}=\tens{\cal{C}}^{(l,l')}\vec{x}
\end{equation}
The covariance matrix $\tens{V}^{(l,l')}_{m,m'}$ of 
$\vec{z}_{\vec{y}}$ ($\vec{z}_{\vec{y}}^{\rm 
T}=(\mbox{Re}(\vec{y}^{\rm T}),\mbox{Im}(\vec{y}^{\rm T}))$), 
can be derived from the sub-matrix $\tens{\cal{V}}^{(l,l')}$ whose 
elements can be expressed as:
\begin{eqnarray}
2\tens{\cal{V}}^{(l,l')}_{m,m'}=E[y_{l,m}(\nu)y_{l',m'}^{*}(\nu)]=\nonumber\\
2 \sum_{l"=l,l'}\sum_{m"=-l"}^{m"=l"} \tens{\cal{C}}_{m',m"}^{(l',l")}
\tens{\cal{C}}_{m,m"}^{(l,l")*}f_{m"}^{l''}(\nu)+2\tens{\cal{B}}_{m,m'}^{(l,l')}
\label{imp}
\end{eqnarray}
where $E$ is the expectation, $f_{m''}^{l''}(\nu)$ is the profile of 
the $l'',m''$ mode, 
$\tens{\cal{B}}^{(l,l')}$ is the covariance matrix of the noise, and 
with the $y_{l,m}(\nu)$ having a mean of zero.  The 
factor 2 
comes from the fact that the real part of $\tens{\cal{V}}^{(l,l')}$ 
represents both the covariance of the 
real or imaginary parts of the Fourier spectra (See Part I, section 
3.3.2); the same property applied to the 
imaginary part of $\tens{\cal{V}}^{(l,l')}$ which represents the 
covariance between the real part and the imaginary part of the 
Fourier 
spectra.
Equation (\ref{imp}) contains all the information that we need for 
visualizing an estimate of the real and imaginary parts of 
$\tens{\cal{V}}^{(l,l')}$.  Drawing from the usefulness of the 
diagrammes of Grec and Brown,  we 
created four 
new diagrammes for visualizing an estimate of 
$\tens{\cal{V}}^{(l,l')}$, all having 
various diagnostics power:
\begin{enumerate}
\item $(m,\nu)$ echelle diagramme: estimate of the diagonal elements 
of $\tens{\cal{V}}^{(l,l)}$ ($l=l'$)
\item cross echelle diagramme: estimate of the off-diagonal elements 
of $\tens{\cal{V}}^{(l,l)}$ ($l=l'$)
\item inter echelle diagramme: estimate of the off-diagonal elements 
of $\tens{\cal{V}}^{(l,l')}$ ($l /ne l'$)
\item ratio cross spectrum: estimate of the ratio of the elements of 
$\tens{\cal{B}}^{(l,l')}$
\end{enumerate}
Each diagnostic is described hereafter in more detail.

\subsection{$(m,\nu)$ echelle diagramme}
The \emph{$(m,\nu)$ echelle diagramme} is made of $2l+1$ echelle 
diagrammes of each $l,m$ power spectra or $|y_{l,m}(\nu)|^{2}$.
  The $2l+1$ echelle diagrammes 
are stacked on top of each other to show the dependence  of 
the mode frequency upon $m$.  These diagrammes give an estimate of 
the 
diagonal of the covariance matrix of the observations as:
\begin{equation} 
2\widetilde{\tens{\cal{V}}}_{m,m}^{(l,l)}(\nu)=|y_{l,m}(\nu)|^{2} 
\end{equation}
where $\widetilde{\tens{\cal{V}}}^{(l,l)}$ symbolizes the estimate of 
$\tens{\cal{V}}^{(l,l)}$.  It is important when one makes these 
diagrammes to tune the spacing for 
the degree to study.  The spacing for a given $l$ can be computed 
from available p-mode frequencies.  Since the spacing varies with the 
degree, other 
modes 
with a significant different spacing can be seen more like diagonal 
ridges crossing the 
$(m,\nu)$ diagrammes; this is a powerful tool to identify 
other degrees.    

Nevertheless, the diagnostics power of the $m,\nu$ 
echelle diagramme is rather limited for deriving the leakage matrix: 
it 
can be shown using Eq. (\ref{imp}) that the diagonal elements of 
$\tens{\cal{V}}^{(l,l)}$ 
can be expressed as:
\begin{eqnarray}
\tens{\cal{V}}^{(l,l)}_{m,m}=\sum_{m"=-l}^{m"=l} 
|\tens{\cal{C}}_{m,m"}^{(l,l)}|^{2}f_{m"}^{l}(\nu)+\tens{\cal{B}}_{m,m}^{(l,l)}
\label{imp3}
\end{eqnarray}  
As we can see with Eq. (\ref{imp3}), the sign information 
of the elements of $\tens{\cal{C}}^{(l,l)}$ is lost; second, their 
magnitude being 
typically less than 0.5, the leakage elements cannot be easily seen 
in the power spectra.  
Another kind of diagramme that preserves the sign of the leakage 
elements had to be devised.

\subsection{Cross echelle diagramme}
The \emph{cross echelle diagramme} of an $l,m$ mode is made of $2l+1$ 
echelle 
diagrammes of the cross spectrum of 
$m$ and $m'$ 
or $y_{l,m}(\nu)y_{l,m'}^{*}(\nu)$.  
The $2l+1$ real (or imaginary) parts of the cross spectra are stacked 
on 
top of each other to show the 
dependence upon $m$ of 
the mode frequency.   These diagrammes give an estimate of the rows 
(or columns) of the covariance matrix of the observations as: 
\begin{equation} 
2{\widetilde{\tens{\cal{V}}}_{m,m'}^{(l,l)}(\nu)}=y_{l,m}(\nu)y_{l,m'}^{*}(\nu)
\end{equation}
Of course when 
$m=m'$ we get the echelle diagrammes of the previous section.  Only 
$l+1$ cross echelle 
diagrammes are shown as the matrix $\tens{\cal{V}}^{(l,l')}$ is 
hermitian by definition. 

The imaginary part of the cross spectra has some 
diagnostic power: it represents the correlation between the real and 
imaginary parts of the Fourier spectra.  When the leakage 
matrix is real, which is generally the case, there is no correlation 
between the real and 
imaginary parts.  Nevertheless the imaginary part could be helpful to 
find errors in 
the filters applied to the images (See Part I, section 3.3.1).  

It can be shown that the 
elements of $\tens{\cal{V}}^{(l,l)}$ can be expressed as:
\begin{eqnarray}
\tens{\cal{V}}_{m,m'}^{(l,l)}=\tens{\cal{C}}_{m,m'}^{(l,l)}f_{m'}^{l}(\nu)+\tens{\cal{C}}_{m',m}^{(l,l)*}
f_{m}^{l}(\nu)+\nonumber\\
\sum_{m" \ne m',m}\tens{\cal{C}}_{m',m"}^{(l,l)}
\tens{\cal{C}}_{m,m"}^{(l,l)*}f_{m''}^{l''}(\nu)+\tens{\cal{B}}_{m,m'}^{(l,l)}
\label{imp1}
\end{eqnarray}
As we can see with Eq. (\ref{imp1}), these diagrammes preserve 
the sign of the leakage matrix elements.  
In general, the cross spectra for $m,m'$, 
representing $\tens{\cal{V}}_{m,m'}^{(l,l)}$ carries 
information over the sign of the leakage elements 
$\tens{\cal{C}}^{(l,l)}_{m,m'}$ 
and $\tens{\cal{C}}^{(l,l)*}_{m',m}$.  The other additional terms 
expressed as product of 
leakage elements are sometimes more difficult to interpret.

But the power of these diagrammes is not only restricted to checking 
the 
sign of the elements of the leakage matrix.
They are also real tools to get a first order estimate of the leakage 
matrix.  We have shown in Part I, that there is
no difference between fitting data for which the leakage matrix 
\emph{is} the identity, and data for which 
the leakage matrix is \emph{not}.  We showed in part I, that the 
covariance matrix of $\vec{\tilde{x}}$ 
($\vec{\tilde{x}}=\tens{\cal{C}}^{-1}\vec{y}$) can be written, 
similarly to that of Eq. (\ref{imp})), by 
the sum of 2 matrices: the first one represents the mode covariance 
matrix and is diagonal, and the second term represents the covariance 
matrix of the noise.  Therefore, applying the inverse of the leakage 
matrix to the data should, in principle, remove all artificial mode
correlations between the Fourier spectra of $\vec{\tilde{x}}$: this 
can be verified using 
the cross echelle diagrammes.  This is the most 
powerful test for deriving the leakage matrices.

The cross echelle diagrammes are useful to verify the correlation 
within a given 
degree, but other degrees are known to leak into the target degree, 
such as $l=6$, 7 
and 8 into $l=1$, 4 and 8, respectively.  The purpose of the next 
diagramme is to assess the magnitude of these leakages.

\subsection{Inter echelle diagramme}
The \emph{inter echelle diagramme} of an $l,m$ mode for the degree 
$l'$ is made of $2l'+1$ echelle 
diagrammes of the cross spectrum of $l,m$ and 
$l',m'$ or $y_{l,m}(\nu)y_{l',m'}^{*}(\nu)$.  The $2l'+1$ real part 
of 
the cross 
spectra are stacked on top of each other to show the dependence upon 
$m'$.
These diagrammes give an estimate of the rows (or columns) of the 
covariance matrix of the observations as: 
\begin{equation} 
2{\widetilde{\tens{\cal{V}}}_{m,m'}(\nu)}=y_{l,m}(\nu)y_{l',m'}^{*}(\nu)
\end{equation}
Similarly as for the cross echelle diagramme, it will help 
to visualize the covariance matrix between different degrees, and to 
derive leakage elements of the full leakage matrix 
$\tens{\cal{C}}^{(l,l')}$.  One can derive an equation similar to Eq. 
(\ref{imp1}) 
for different degrees showing that the inter echelle diagramme 
carries 
information over the sign of the leakage elements 
$\tens{\cal{C}}^{(l,l')}_{m,m'}$ 
and $\tens{\cal{C}}^{(l',l)*}_{m',m}$.

As mentioned above 
applying the inverse of the leakage matrix will help to verify to the 
first order that there is no artificial correlation due to the p 
modes.  When different degrees are 
involved the full leakage matrix $\tens{\cal{C}}^{(l,l')}$ has to be 
used, 
producing diagrammes that should have no artificial correlation due 
to the p modes.

\subsection{Ratio cross spectrum}
All the previous diagrammes are helpful to understand and visualize 
the mode covariance matrices.  Unfortunately, due to the high 
signal-to-noise ratio, these diagrammes cannot be used to evaluate 
the correlation of the noise in the Fourier spectra.  In between the 
p modes, these correlations can be more 
easily visualized as we have:
\begin{eqnarray}
\tens{\cal{V}}_{m,m'}^{(l,l)}\approx\tens{\cal{B}}_{m,m'}^{(l,l)}
\label{imp2}
\end{eqnarray}
But instead of visualizing $\tens{\cal{B}}^{(l,l')}$, we prefer to 
look 
directly at the correlation by computing the ratio of the cross 
spectra 
as:
\begin{equation}
\widetilde{\frac{\tens{\cal{B}}^{(l,l')}_{m,m'}}{\tens{\cal{B}}^{(l,l)}_{m,m}}}=
\frac{(y_{l,m}(\nu)y_{l,m'}^{*}(\nu)}{|y_{l,m}(\nu)|^{2}}
\end{equation}
This ratio is called the \emph{ratio cross spectrum}.    The ratio 
cross 
spectrum gives an estimate of the ratio matrix 
$\tens{\cal{R}}^{(l,l')}$ 
(See Part I) which gives 
a better understanding of how much the noise background is correlated 
between the Fou-rier spectra.  Nevertheless, in the p-mode frequency 
range, the
 ratio cross spectrum is more difficult to interpret as
the noise correlation is affected by the presence of the modes.  
By looking away from the modes (at high or low 
frequency) or by looking between
the modes, one could obtain a reasonable good estimate of the noise 
correlation. 

\begin{figure*}[!]
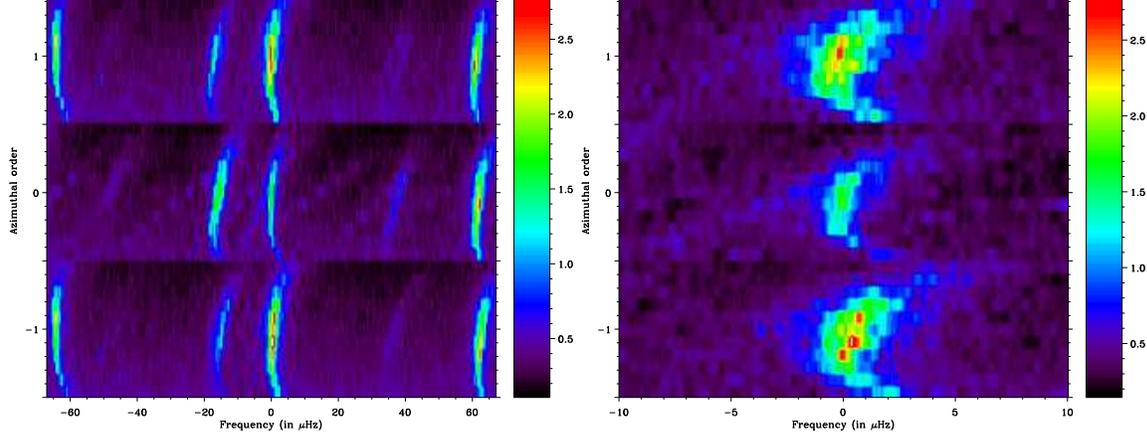

\centering
\includegraphics[angle=90,width=7.5cm]{echelle1.epsi}
\includegraphics[angle=90,width=7.5cm]{echelle1fine.epsi}
\caption{$(m,\nu)$ echelle diagramme for 1 year of LOI data for 
$l=1$.  The scale is in ppm\,$\mu$Hz$^{-1/2}$. The spacing is tuned 
for $l=1$.  (Left) The full diagramme centered on $l=1$.  The $l=3$ 
modes are located about 15 $\mu$Hz at the 
left hand side of the $l=1$.  The $l=2$ modes are on the right edge, 
while the 
$l=0$ are on the opposite side.  The $l=4$ modes can 
be seen around +40 $\mu$Hz. (Right) The same diagramme but enlarged 
around $l=1$.  The frequency shift or splitting of the 
modes due to the solar rotation can be seen: the 2 patches of power 
for $m=-1$ and $m=+1$ are slightly displaced from each other.  The 
artificial 
correlation between $m=-1$ and $m=+1$ is not as clear.}
\label{mnuLOI1} 
\end{figure*}

\begin{figure*}[!]
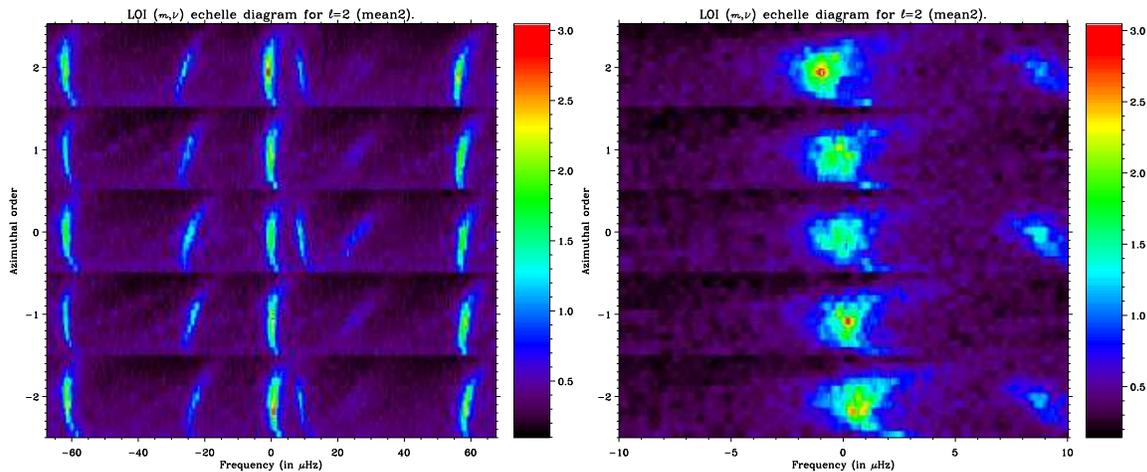

\centering
\centerline{\hbox{
\includegraphics[angle=90,width=7.5cm]{echelle2.epsi}
\includegraphics[angle=90,width=7.5cm]{echelle2fine.epsi}
}}
\caption{$(m,\nu)$ echelle diagramme for 1 year of LOI data for 
$l=2$. The scale is in ppm\,$\mu$Hz$^{-1/2}$.  The spacing is tuned 
for $l=2$.  (Left) The full diagramme centered on $l=2$ with a 
spacing of 
135. $\mu$Hz.  The $l=0$ modes are located about 10 $\mu$Hz at the 
right hand side of the $l=2$ modes.  The $l=3$ modes are on the right 
edge, while the 
$l=1$ modes are on the opposite side.  The $l=4$ modes are easily 
seen at 
about -25 $\mu$Hz; the $l=5$ starts to appear at +25 $\mu$Hz. (Right) 
The same diagramme but enlarged 
around $l=2$ with the $l=0$ modes at the right hand side.  The 
splitting of 
the 
modes is clear.  Note the absence of the $l=0$ modes for 
$m=\pm 
1$}
\label{mnuLOI2} 
\end{figure*}

\begin{figure*}[!]
\centerline{\hbox{
\includegraphics[angle=90,width=7.5cm]{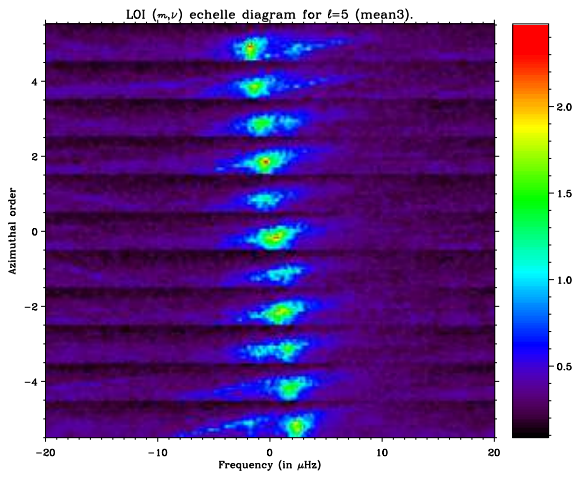}}
\hbox{
\hfill
\parbox[b]{87mm}{
\caption{$(m,\nu)$ echelle diagramme for 1 year of LOI data for 
$l=5$.  The scale is in ppm\,$\mu$Hz$^{-1/2}$.  The spacing is tuned 
for $l=5$.  Here we can clearly see the 
splitting of the $l=5$ modes.  Unfortunately, the $l=5$ data of the 
LOI are heavily polluted by the presence of the $l=8$ modes.  But 
since the 
spacing for the $l=8$ modes is different compared with that of the 
$l=5$ modes, the 
$l=8$ modes are 
seen as ridges going from lower left to upper right in each $m$ 
panel; while the 
$l=5$ modes are seen more like straight ridges in each $m$ panel.  It 
can also be 
seen that the $|m|=5$ Fourier spectra contain information both about 
$m=\pm 5$.  This is the result of the large LOI pixel size creating 
oversampling.  The $l=8$ modes are also distinguished because the 
`splitting' of this alias is in the opposite direction compared with 
that of the 
$l=5$ modes.}
\label{mnuLOI5}}
}} 
\end{figure*}

\vspace{0.5 cm}

\section{Application to data}
\subsection{($m,\nu$) echelle diagramme for the LOI/SOHO data}
Example of these diagrammes can be seen in Figs. 
\ref{mnuLOI1}, \ref{mnuLOI2} and \ref{mnuLOI5} for 1 year of LOI data 
for $l=1,2$ and 5, respectively.  It 
is important when one makes such diagrammes to tune the spacing for 
the degree to study.  For example, one can see in Fig. \ref{mnuLOI1} 
that the ridges of power of $l=0,1,2$ 
and 3 have different shapes than in Fig. \ref{mnuLOI2}.  Other modes 
with a significant 
different spacing can be seen more like diagonal ridges crossing the 
$(m,\nu)$ diagrammes; this is a powerful tool to identify 
other degrees.  In Fig. \ref{mnuLOI5}, the 
$l=5$ 
$(m,\nu)$ echelle diagramme is clearly contaminated by an other 
degree, i.e. $l=8$, which appears at different frequencies depending 
on $m$.  For 
$l=5, 
m=-5$, the $l=8, m=+8$ is quite strong; while for $l=5, m=+5$, this 
is $l=8, m=-8$ which shows up.  This kind of `anti'-splitting 
behaviour is typical of any aliasing degrees.  It is more prominent 
in the LOI data because of the undersampling effect due to the large 
size of the LOI pixels.

\begin{figure*}[!]
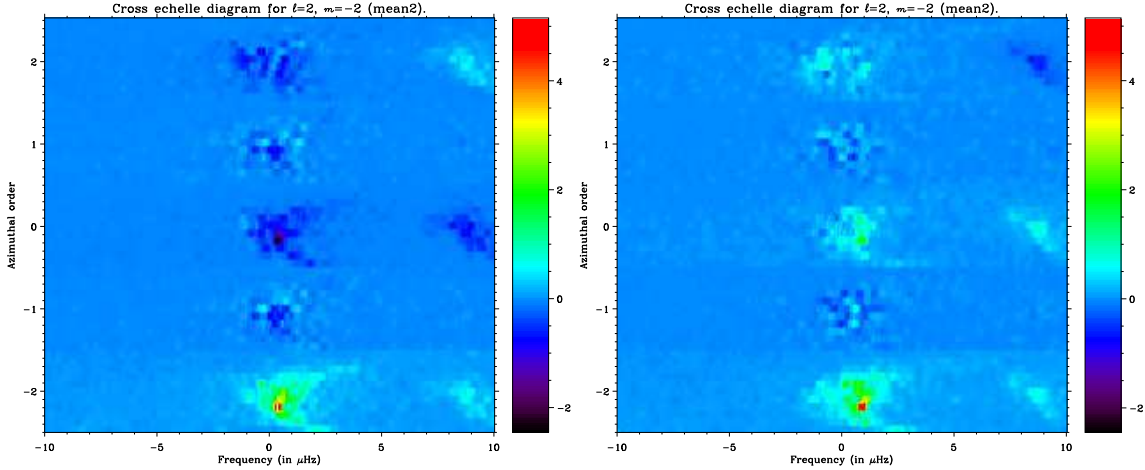

\centerline{\hbox{
\includegraphics[angle=90,width=7.5cm]{echelle22r.epsi}
\includegraphics[angle=90,width=7.5cm]{echelle22i.epsi}
}}
\caption{Effect of a wrong sign on the cross echelle diagrammes for 1 
year of LOI data for $l=2, m=-2$.  The scale is in 
ppm$^{2}$\,$\mu$Hz$^{-1}$.  The spacing is tuned for $l=2$. 
(Left) For the real parts or 
$\mbox{Re}(y_{l,m}(\nu))\mbox{Re}(y_{l,m'}(\nu))$. (Right) For the 
imaginary 
parts or $\mbox{Im}(y_{l,m}(\nu))\mbox{Im}(y_{l,m'}(\nu))$ without 
having the sign corrected.  
The real part of the cross echelle diagramme is constructed by the 
sum of 
those 2 
quantities; compensation due to different signs 
will be misleading.  The correlation of the real parts will always 
have the correct sign. This can be used for checking that the 
correlation of the imaginary parts for $m=0$ and $m=2$ is different 
from that of the correlation of the real parts.}
\label{error2} 
\end{figure*}

\begin{table*}[!]
\caption{Theoretical leakage matrix of $l$=5 for the LOI data.}
\label{c55}
\begin{center}
\begin{tabular}{ccccccccccccc}
\hline 
&& & & & & &$m'$& & & & \\
&&-5 &	-4 & -3 & -2 &  -1 & 0 & 1 & 2&3&4&5\\  
\hline 
 &-5&     1 &  0  &      0.18& 0&    -0.10&  0& 0.08&0  &    0.19&   
0&     -0.45\\
  &-4&     0 &   1&  0&     0.27&  0&     -0.13&0&      0.17&   
0&     0.03&       0\\
 &-3&     0.18&       0&       1&       0&      0.64&       0& 
-0.21&   0&     -0.62&       0&      0.19\\
  &-2&0&     0.27&  0&       1&0&      0.50&0&     -0.18&   0&      
0.17&0\\
  & -1& -0.10&       0&      0.64&       0&1&       0&0.52&       
0&     -0.21&       0&     0.08\\
$m$&    0&   0&-0.13&0&      0.50&  0&1&0&     0.50&0&-0.13&0\\
    &1& 0.08&0&  -0.21&      0&      0.52&       0&1&       0& 
0.64&    0&   -0.10\\
   &2&0&      0.17&  0&     -0.18&  0&     0.50&   0&     1&  0&     
0.27&  0\\
    &3&  0.19&       0&-0.62&0&     -0.21&      0&0.64&0& 1&     
0&     0.18\\
    &4&   0&     0.03&  0&    0.17&  0&    -0.13&  0&      0.27&  0  
&   1&       0\\
    &5& -0.45&   0&     0.19&  0&     0.08&      0&    -0.10&   0&    
0.18&0  &     1\\
\hline 
\end{tabular} 
\end{center} 
\end{table*}

\vspace{0.5cm}

\subsection{A useful detail}
Before using the other diagrammes on real data, we need to point out 
a 
very important property coming from the way the $m$ signals are built.
If the weights $W_{l,m}$ applied to the images, to extract an $l,m$ 
mode, have the same properties as of 
the spherical harmonics then we have:
\begin{equation}
W_{l,m}=W_{l,-m}^{*}
\end{equation}
which means that both the Fourier spectra of $+m$ and $-m$ can be 
obtained 
from 
using a single filter $W_{l,+m}$. The Fourier spectrum of $+m$ will 
be 
in the 
positive part of 
the frequency range, while that of $-m$ will be in the 
negative part.  This approach was first used by Rhodes \etal (1979) 
for measuring solar rotation, and mentioned theoretically by 
Appourchaux and Andersen (1990) for the case of the LOI.  Due to the 
property of the Fourier transform, we 
recover, in the negative part of the frequency range, \emph{not} the 
Fourier spectrum of $-m$ but its \emph{conjugate}.  This fact is very 
important, if one does not 
take care of the sign of the imaginary part of $-m$, it will lead to 
very serious problem.  Needless to say that fitting data without 
taking into 
account this detail will have devastating effects.  Figure 
\ref{error2} shows 2 cross echelle 
diagrammes for the real and imaginary parts of the spectra of $l=2$; 
the 
latter part had no sign compensation for $m<0$.  Obviously this 
important detail cannot be detected in the power spectra.

\subsection{Leakage matrix measurement for a single degree}
\subsubsection{LOI/SOHO data}
According to theoretical computation of the p-mode sensitivities 
of the LOI pixels (Appourchaux and Andersen, 
1990) and using the real shape of the LOI pixels (Appourchaux and 
Telljohann, 1996), the leakage matrices of $l=1$ and 2 are given by:
\begin{eqnarray}
\tens{\cal{C}}^{(1,1)}=\left(\begin{array}{ccc}
1 & 0 & \alpha\\
0 & 1 & 0 \\
\alpha & 0 & 1\\
\end{array}\right)\nonumber\\
\tens{\cal{C}}^{(2,2)}=\left(\begin{array}{ccccc}
1 & 0 & \alpha_{1}& 0 & \alpha_{4}\\
0 & 1 & 0 & \alpha_{2} & 0\\
\alpha_{3} & 0 & 1 & 0 & \alpha_{3}\\
0 & \alpha_{2} & 0 & 1 & 0\\
\alpha_{4} & 0 & \alpha_{1} & 0 & 1\\
\end{array}\right)
\label{leakageLOI}
\end{eqnarray}
with: $\alpha=0.474$, $\alpha_{1}=\alpha_{3}=-0.308$, 
$\alpha_{2}=0.576$, $\alpha_{4}=-0.216$.  
The leakage matrix of $l=5$ is given in Table \ref{c55}.  These 
leakage matrices are mean value over one year.  
The leakages vary 
throughout the year because the distance between SOHO and the Sun 
varies. There is no $B$ angle effect as the mean $B$ 
is zero over 1 year.  All the leakage elements are real as the LOI 
filters have the same
symmetry as the spherical harmonics.

\begin{figure*}[t]
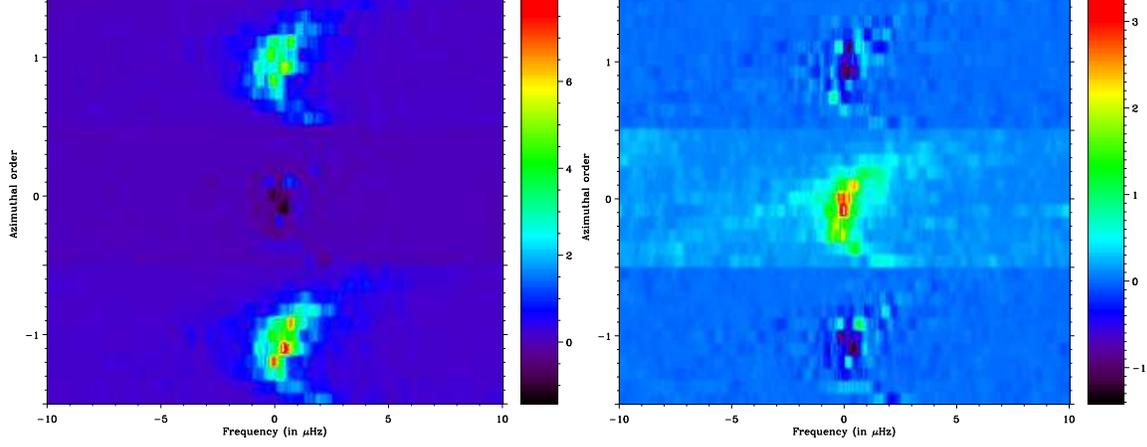

\centerline{\hbox{
\includegraphics[angle=90,width=7.5cm]{echelle11t.epsi}
\includegraphics[angle=90,width=7.5cm]{echelle10t.epsi}
}}
\caption{Real part of the cross echelle diagrammes for 1 year of LOI 
data for $l=1$.  
The scale is in ppm$^{2}$\,$\mu$Hz$^{-1}$.  The spacing is tuned for 
$l=1$.  
(Left) For $l=1, m=-1$.  The lower panel represents the echelle 
diagramme of $m=-1$.  The 2 other panels are the cross spectra with 
$m=0$ and 
$m=1$.  As predicted, there is no visible correlation between $m=-1$ 
and $m=0$ 
while there is between $m=-1$ and $m=+1$.  Here we visualize the 
first line of the real part of the matrix $\tens{\cal{V}}^{(1,1)}$.  
(Right) For $l=1, m=0$.  The bottom 
panel is the same as the middle panel of the left diagramme but with 
a 
different color scale.  The middle panel is the power spectra of 
$m=0$ 
already shown in Fig. 2; the $l=6$ modes are visible as a diagonal 
ridge crossing the ridge of the $l=1$ modes.}
\label{cross1} 
\end{figure*}

\begin{figure*}[!]
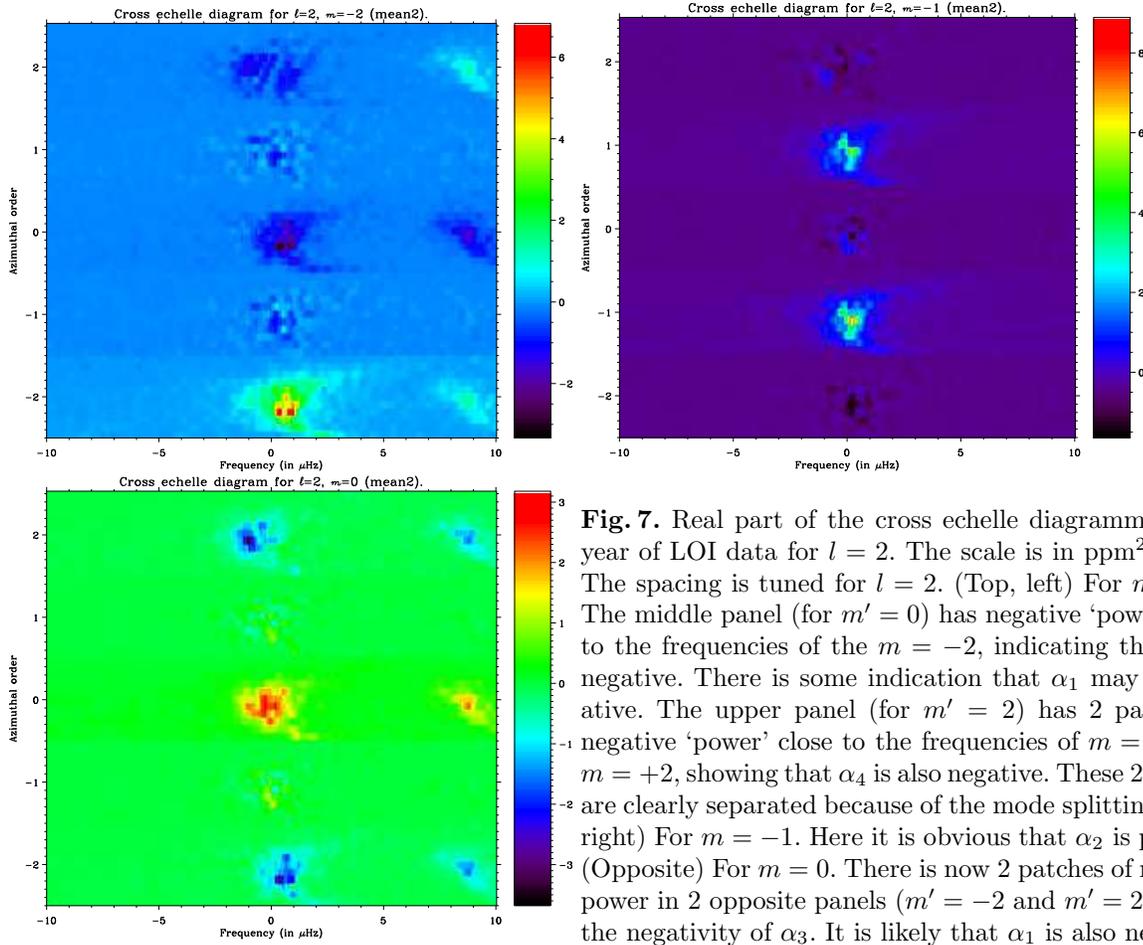

\centerline{\vbox{
\hbox{
\includegraphics[angle=90,width=7.5cm]{echelle22t.epsi}
\includegraphics[angle=90,width=7.5cm]{echelle21t.epsi}
}
\hbox{
\includegraphics[angle=90,width=7.5cm]{echelle20t.epsi}
\hfill
\parbox[b]{87mm}{
\caption{Real part of the cross echelle diagrammes for 1 year of LOI 
data for $l=2$.  
The scale is in ppm$^{2}$\,$\mu$Hz$^{-1}$.  The spacing is tuned for 
$l=2$.   
(Top, left) For $m=-2$.  The middle panel (for $m'=0$) has negative 
`power' 
close to the frequencies of the $m=-2$, indicating that $\alpha_{3}$ 
is negative.  There is some indication that $\alpha_{1}$ may be 
negative.  The upper panel 
(for $m'=2$) has 2 patches of negative `power' close to the 
frequencies of  $m=-2$ and $m=+2$, showing that $\alpha_{4}$ is also 
negative.  These 2 patches are clearly separated because of the mode 
splitting.  (Top, right) For $m=-1$.  Here it is obvious that 
$\alpha_{2}$ is positive. (Opposite) For $m=0$.  There is now 2 
patches of negative 
power in 2 opposite panels ($m'=-2$ and $m'=2$) due to the negativity 
of $\alpha_{3}$.  It is likely that $\alpha_{1}$ is also negative.}
\label{cross2}} 
}}}
\end{figure*}

\begin{figure*}[p]
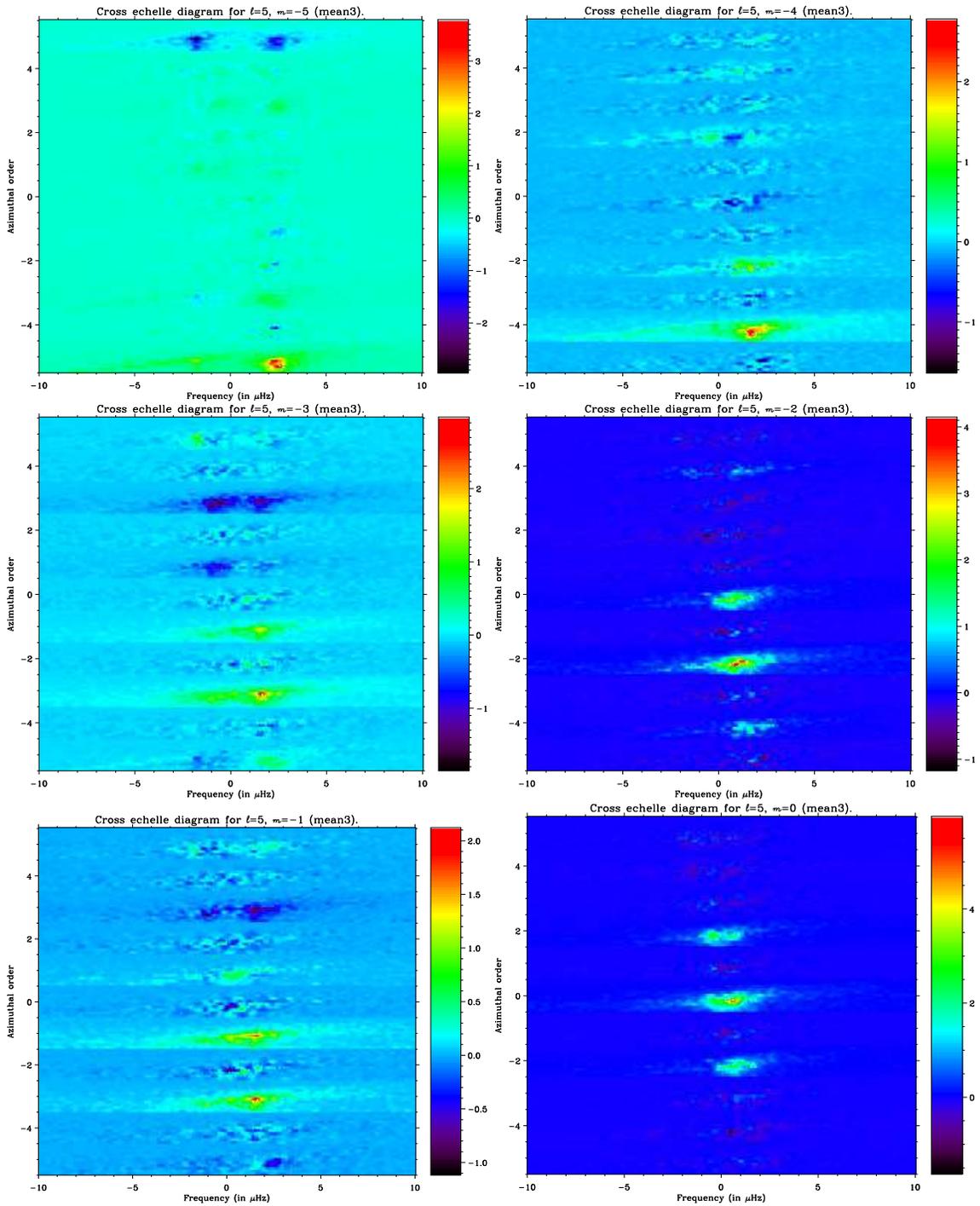

\centerline{\vbox{
\hbox{
\includegraphics[angle=90,width=7.5cm]{echelle55t.epsi}
\includegraphics[angle=90,width=7.5cm]{echelle54t.epsi}}
\hbox{
\includegraphics[angle=90,width=7.5cm]{echelle53t.epsi}
\includegraphics[angle=90,width=7.5cm]{echelle52t.epsi}}
\hbox{
\includegraphics[angle=90,width=7.5cm]{echelle51t.epsi}
\includegraphics[angle=90,width=7.5cm]{echelle50t.epsi}}
}}
\caption{Real part of the cross echelle diagrammes for 1 year of LOI 
data for $l=5$.  
  The scale is in ppm$^{2}$\,$\mu$Hz$^{-1}$.  The spacing is tuned 
for $l=5$. 
(Top, left) For $m=-5$. Here we can see the strong leakage of 
$m'=5$ into $m=5$ and vice versa.  The 2 patches of negative `power' 
on the upper panel (or positive on the lower panel) are the $m=\pm 5$ 
modes separated by about 4-5 $\mu$Hz. (Top, right) For $m=-4$.  
Leakages between $m=-4$ and $m=-2$ are visible, but weaker than for 
$m=-5$.  (Middle, left) For $m=-3$.  Strong leakages between $m=-3$ 
and $m=+3$.  (Middle, right) For $m=-2$. (Bottom, left) For $m=-1$.  
(Bottom, right) For $m=0$.  The left diagrammes show more leakages 
compared with the right ones, the $l+m$ parity makes the difference.}
\label{cross5} 
\end{figure*}

\begin{figure*}[t]
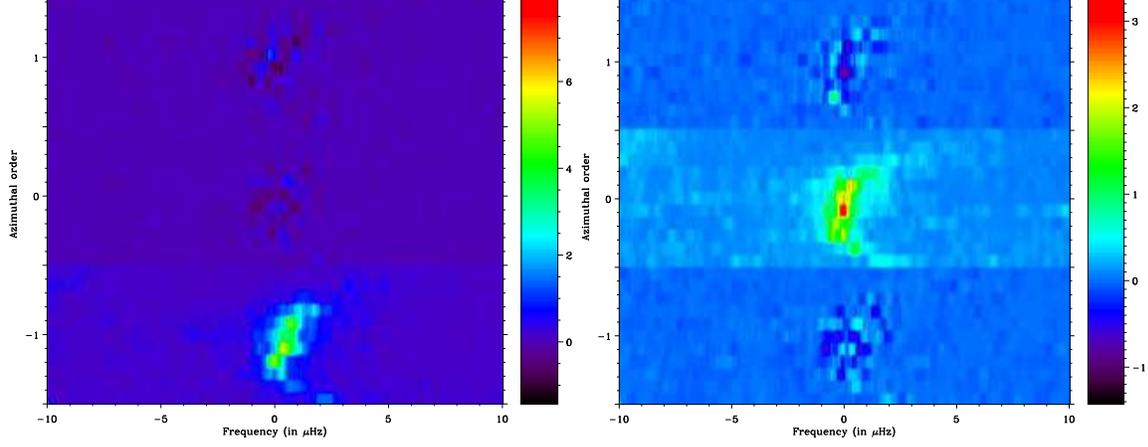

\centerline{\hbox{
\includegraphics[angle=90,width=7.5cm]{year11t.epsi}
\includegraphics[angle=90,width=7.5cm]{year10t.epsi}
}}
\caption{Real part of the cross echelle diagrammes of 1 year of LOI 
data for $l=1$.  
The scale is in ppm$^{2}$$\mu$Hz$^{-1}$.
The inverse of the theoretical
leakage matrix has been applied to the original data with 
$\alpha$=0.474.
(Left) For $l=1, m=-1$.  The artificial correlation between $m=-1$ 
and $m=+1$ has been
entirely removed (See Fig. \ref{cross1} for comparison).  (Right) For 
$l=1, m=0$.  There is no improvement 
as there was no correlation before.}
\label{year1} 
\end{figure*}

\begin{figure*}[!]
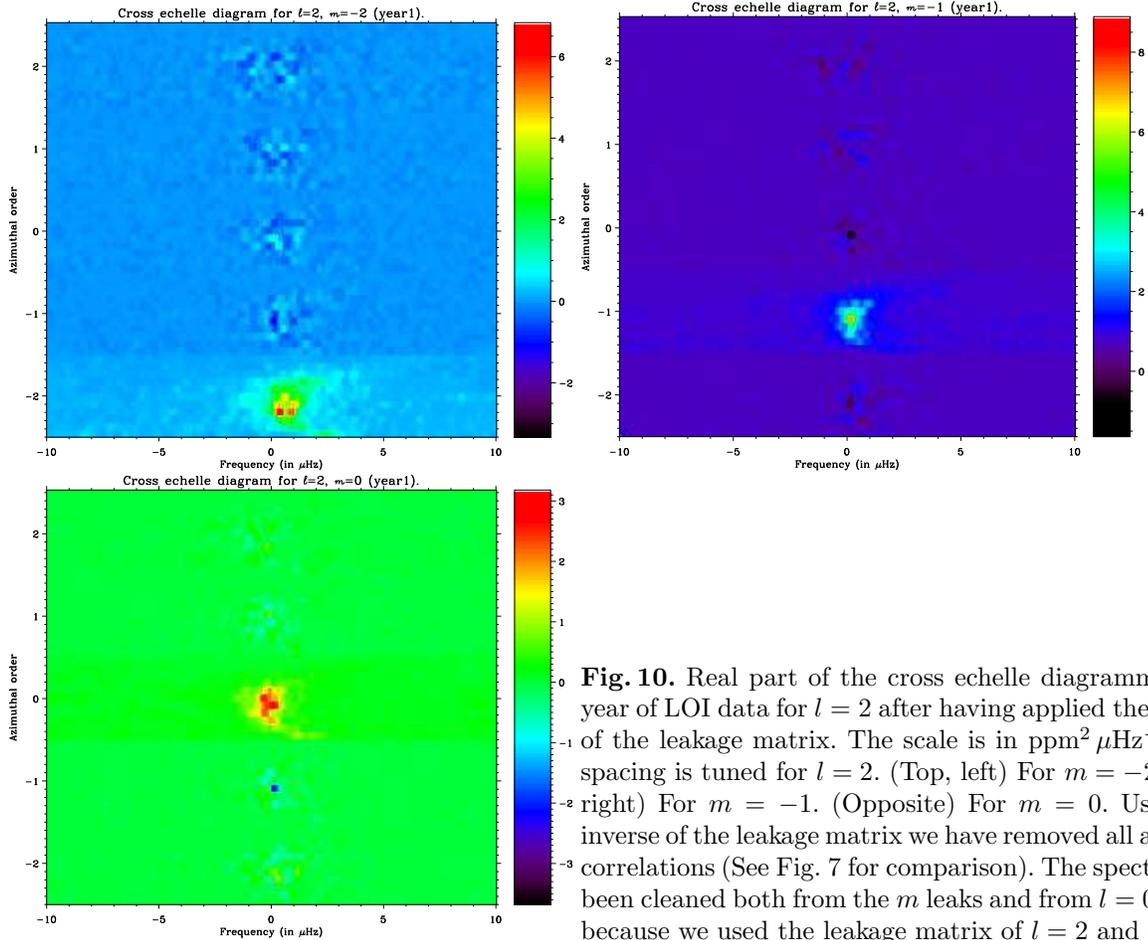

\centerline{\vbox{
\hbox{
\includegraphics[angle=90,width=7.5cm]{year22t.epsi}
\includegraphics[angle=90,width=7.5cm]{year21t.epsi}}
\hbox{
\includegraphics[angle=90,width=7.5cm]{year20t.epsi}
\hfill
\parbox[b]{87mm}{
\caption{Real part of the cross echelle diagrammes of 1 year of LOI 
data for $l=2$ 
after having applied the inverse of the leakage matrix.    The scale 
is in ppm$^{2}$\,$\mu$Hz$^{-1}$.  The spacing is tuned for $l=2$.
(Top, left) For $m=-2$.  (Top, right) For $m=-1$. (Opposite) For 
$m=0$.  Using the inverse of the leakage matrix we have removed all 
artificial correlations (See Fig. \ref{cross2} for comparison).  The 
spectra have been cleaned both from 
the $m$ leaks and from $l=0$ modes because we used the leakage 
matrix of $l=2$ and $l=0$.}
\label{year2}} 
}}}
\end{figure*}

Figure \ref{cross1}, \ref{cross2} and \ref{cross5} display the cross 
echelle diagramme for $l=1,2$ and 5, respectively.  
From Figs. \ref{cross1} and \ref{cross2}, we can 
directly verify using Eq. (\ref{imp1}) the sign, and sometimes even 
the 
magnitude of 
the leakage elements.  The 
best cross check that 
our theoretical computations are correct
is to apply the inverse of the leakage matrices to the original 
data (See Section 2.2).  Figures \ref{year1} and \ref{year2} show how
the artificial correlations (or $m$ leaks) can be removed from the 
LOI data; for the 
latter we 
also cleaned the original data from the
presence of the $l=0$ modes.  We shall see later on with the GONG 
data 
that cleaning, 
similar to that of the LOI, can be achieved not 
only for 2 degrees but also for 3 degrees ($l=1,6$ and 9).

Unfortunately, the leakage matrix of 
$l=5$ for the LOI data cannot 
be inverted.  This is the result of the pixel undersampling and has 
two dramatic consequences: the leakage matrix cannot be 
verified as for the $l=2$, and fitting the data as described in 
Part I is not valid anymore as the leakage matrix needs to be 
invertible.  It means that the $2l+1$ Fourier spectra of $l=5$ are 
dependent.  One way around the problem is to restrict the fitting to 
truly 
independent Fourier spectra.  This problem is specific to the LOI 
data and applies only for $l \ge 4$.

\vspace{0.5 cm}

\subsubsection{GONG data}
The leakage matrices of the GONG data have been computed by R.Howe 
(1996, private communication).
We also computed similar leakage matrices using the equations 
developed in Part I.  The integration was made in the $\theta,\phi$ 
plane with $\theta_{max}=65.2^{\circ}, \phi_{max}=53.5^{\circ}$ 
without apodization.  We also took into account the 
effect of subtracting the velocity of the Sun seen as a star which 
affects GONG instrument's sensitivity to modes only detected 
by integrated sunlight instrument (mainly the modes for which $l+m$ 
is even).  The theoretical leakage matrices of $l=1$ and 2 
for GONG are also given by Eq. (\ref{leakageLOI}) but with: 
$\alpha=-0.55$, $\alpha_{1}=-0.268$, $\alpha_{2}=0.451$, 
$\alpha_{3}=-0.122$, and $\alpha_{4}=-0.290$.  For $l=1$, the leakage 
between $m=-1$ and $m=+1$ is negative due to 
the subtraction of the mean velocity which affects these modes.
For $l=2$, the leakage between $m=-1$ and $m=+1$ has about the same 
value as for the LOI; these modes are not affected by the 
subtraction.  

\begin{figure*}[t]
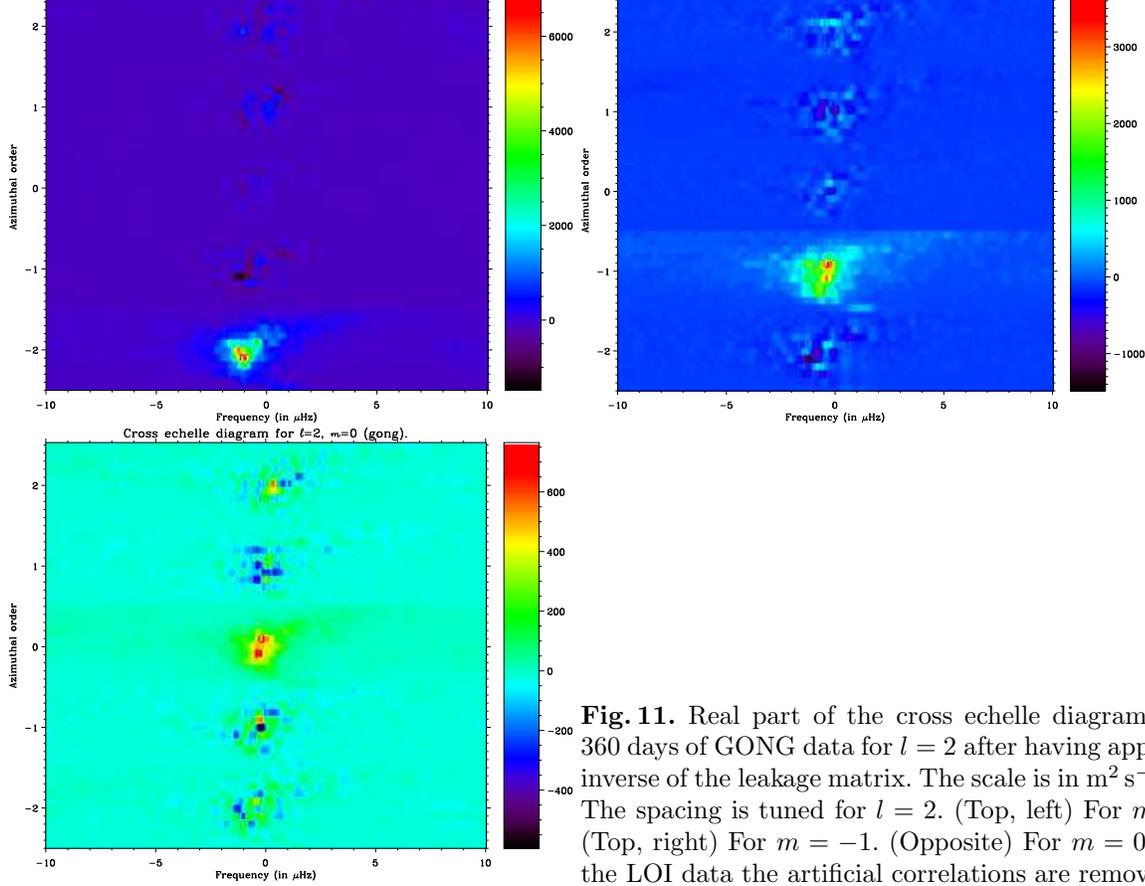

\centerline{\vbox{
\hbox{
\includegraphics[angle=90,width=7.5cm]{gong22t.epsi}
\includegraphics[angle=90,width=7.5cm]{gong21t.epsi}}
\hbox{
\includegraphics[angle=90,width=7.5cm]{gong20t.epsi}
\hfill
\parbox[b]{87mm}{
\caption{Real part of the cross echelle diagrammes for 360 days of 
GONG data for $l=2$ 
after having applied the inverse of the leakage matrix.     The scale 
is in m$^{2}\,$s$^{-2}\,$Hz$^{-1}$.  The spacing is tuned for 
$l=2$.
(Top, left) For $m=-2$.  (Top, right) For $m=-1$. (Opposite) For 
$m=0$.  As for the LOI data the artificial correlations are removed.}
\label{gong2}} 
}}}
\end{figure*}

\begin{figure*}[t]
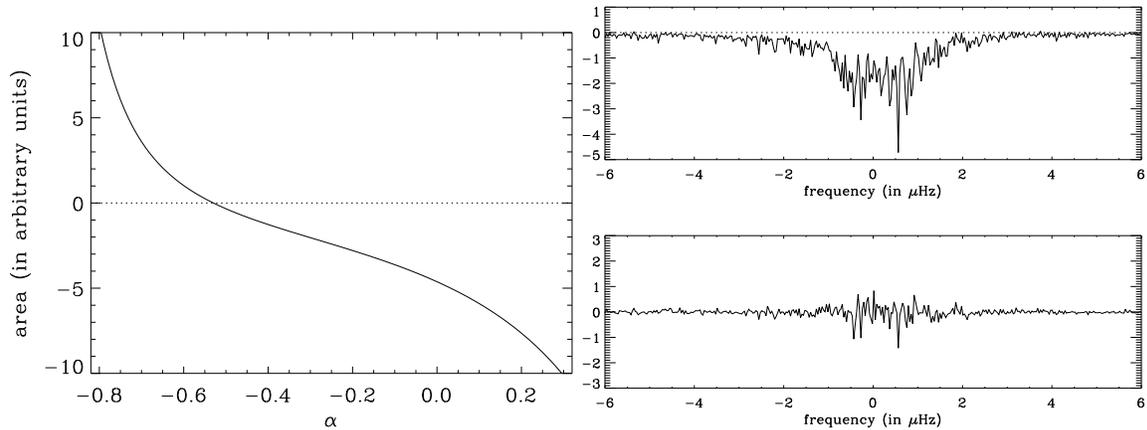

\centerline{\hbox{
\includegraphics[angle=0,width=7.5cm]{area_alphal1.epsi}
\includegraphics[angle=0,width=7.5cm]{colapl1.epsi}}
}
\caption{(Left) Collapsed diagramme of the real part of the cross 
spectrum of $m=-1$ 
and $m=+1$ for the GONG data after having applied the inverse of the 
leakage matrix, (top) no correction, (bottom) $\alpha$=-0.53. (Right) 
Surface of the collapsed diagramme as a function of $\alpha$, the 
surface is 0 for $\alpha$=-0.53.}
\label{alpha}
\end{figure*}

The cross echelle diagrammes of the $l=2$ GONG data are 
very close to those of the LOI (See Fig. \ref{cross2}). Figure 
\ref{gong2} 
shows an example of a GONG cross echelle diagramme after having 
applied the 
inverse of the leakage matrix.  It is clear that the p-mode 
correlations are removed.  For $l=1$ we have used the cross echelle 
diagramme of $m=-1$ and $m=1$ for inferring quantitatively the 
off-diagonal leakage element $\alpha$.  We first applied the inverse 
of a 
leakage matrix to the $l=1$ data, and then constructed the cross 
echelle 
diagramme of $m=-1$ and $m'=1$ for these data.  We then collapsed 
this 
diagramme by adding up all the modes with $n=10-26$, and finally we 
corrected the collapsed diagramme from the solar noise background.  
The collapsed diagramme is shown in Fig. \ref{alpha} (Left) for no 
correction ($\alpha=0$) and for an $\alpha$ of -0.53;  the corrected 
surface of the collapsed diagramme as a function of 
$\alpha$ is shown on Fig. \ref{alpha} (Right).  When the corrected 
surface is close to 0, there is no artificial correlation remaining.

Cleaning the data from artificial correlations has also been done in 
a 
different way by Toutain \etal (1998).  Using Singular Value 
Decomposition, they recomputed pixel filters for the MDI/LOI proxy so 
as 
to remove the $m$ leaks and the other aliasing degrees.  Here we have 
shown, that data 
cleaning is also possible \emph{without} having the pixel time 
series, but using the Fourier spectra.  This latter cleaning 
technique is more useful 
as one has to reduce a smaller amount of data.

\begin{figure*}[!]
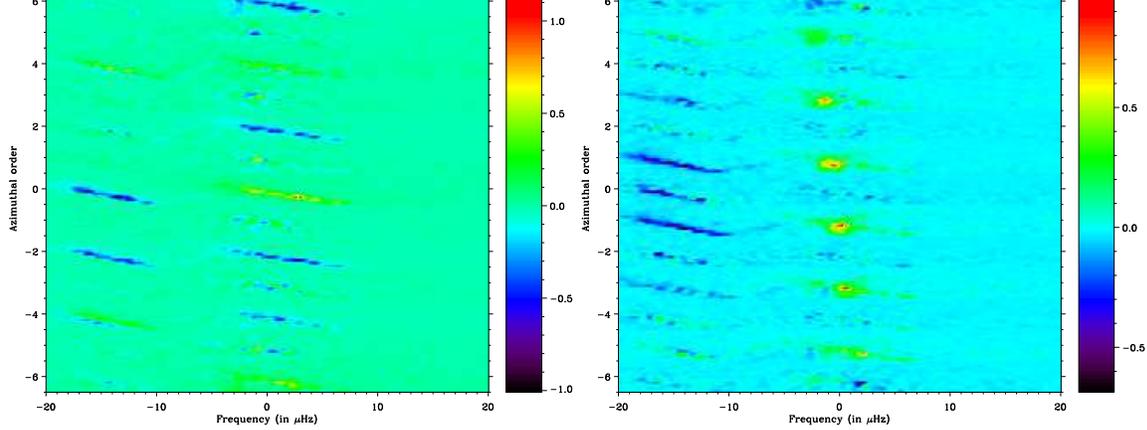

\centerline{\hbox{
\includegraphics[angle=90,width=7.5cm]{echelle161t.epsi}
\includegraphics[angle=90,width=7.5cm]{echelle160t.epsi}}
}
\caption{Real part of the inter echelle diagrammes for 1 year of LOI 
data for $l=1$ 
and $l'=6$. The scale is in ppm$^{2}$\,$\mu$Hz$^{-1}$.  The spacing 
is tuned for $l'=6$.  (Left) For $l=1, m=-1$.  Given the structure of 
the 
ridges, 
it means that the $l=1$ modes leak into the even $m$ modes of $l'=6$, 
while the $l'=6$ modes do not leak into the $l=1$.  (Right) For $l=1, 
m=0$.  Again 
using the 
structure of the ridges, the odd $m'$ of $l'=6$ leaks weakly into 
$l=1, m=0$.}
\label{inter16} 
\end{figure*}

\begin{figure*}[p]
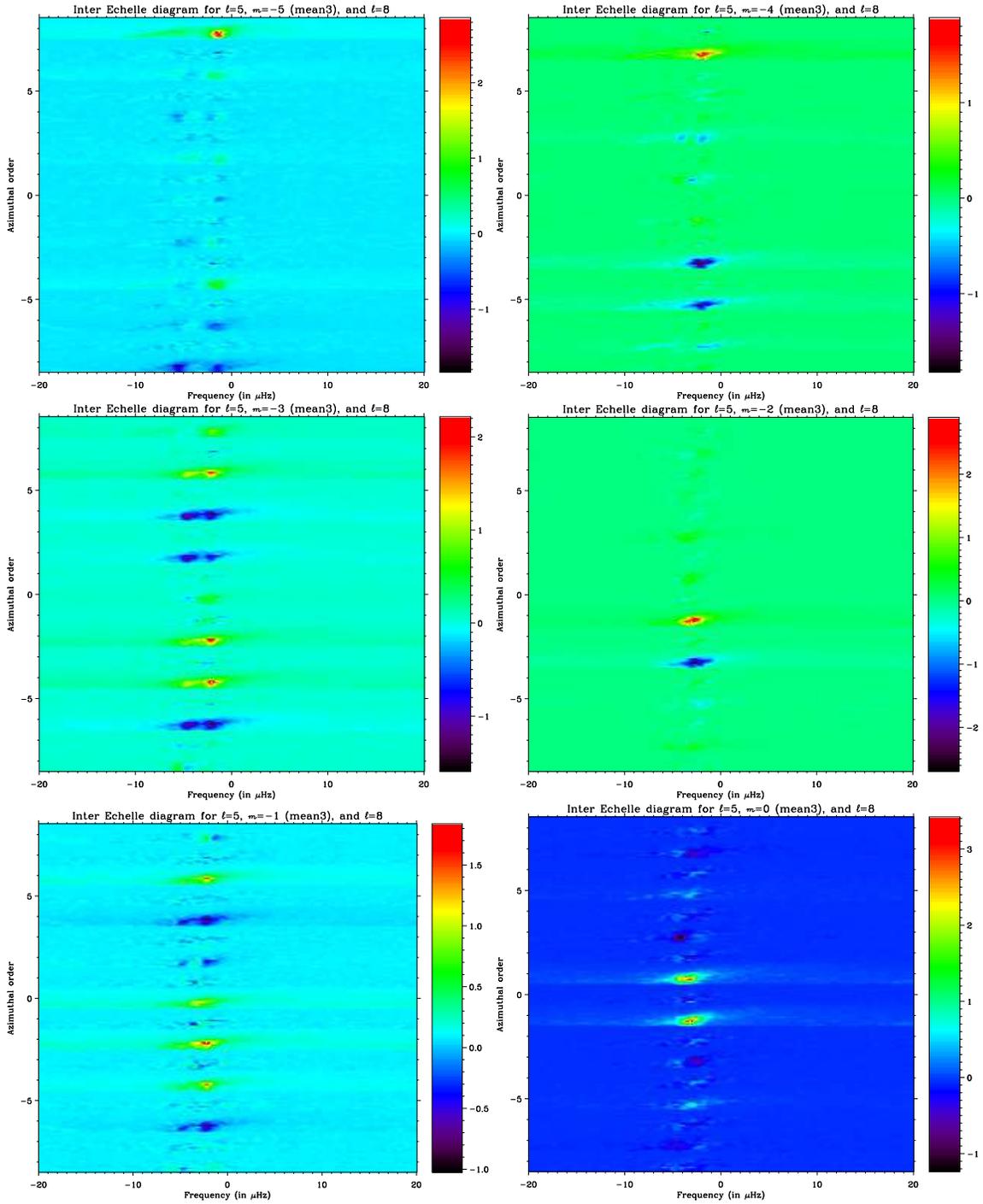

\centerline{\vbox{
\hbox{
\includegraphics[angle=90,width=7.5cm]{echelle585t.epsi}
\includegraphics[angle=90,width=7.5cm]{echelle584t.epsi}}
\hbox{
\includegraphics[angle=90,width=7.5cm]{echelle583t.epsi}
\includegraphics[angle=90,width=7.5cm]{echelle582t.epsi}}
\hbox{
\includegraphics[angle=90,width=7.5cm]{echelle581t.epsi}
\includegraphics[angle=90,width=7.5cm]{echelle580t.epsi}}
}}
\caption{Real part of the inter echelle diagrammes for 1 year of LOI 
data for $l=5$ 
and $l'=8$.  The scale is in ppm$^{2}$\,$\mu$Hz$^{-1}$.  The spacing 
is tuned for $l'=5$.  (Top, left) For $l=5, m=-5$. (Top, right) For 
$l=5, m=-4$. 
(Middle, left) For $l=5, m=-3$. (Middle, right) For $l=5, m=-2$. 
(Bottom, left) For $l=5, m=-1$. (Bottom, right) For $l=5, m=0$.  The 
$l=5$ ridges are more like vertical lines while those of the $l=8$ 
are more like diagonal lines going across from left to right.}
\label{inter58} 
\end{figure*}

\vspace{0.5 cm}

\subsection{Leakage matrix measurement for many degrees}
\subsubsection{LOI/SOHO data}
Figures \ref{inter16} and \ref{inter58} shows the inter echelle 
diagramme for 
$l=1$, $l'=6$, and $l=5$, $l'=8$, respectively.  These 
diagrammes are 
more difficult to interpret, but
using specific spacing for $l$ or $l'$, one can unambiguously 
identify which degree contributes to the correlation.  For the LOI 
data, the 
$l=5$ and 
$l=8$ (similarly $l=4$ and $l=7$) are contaminated by each other.
This is again due to the spatial undersampling which prevents the LOI 
data to be cleaned from aliasing degrees making the leakage matrix 
not 
invertible.  Fortunately, the $l=1$ LOI data 
are far less affected by the presence of $l=6$ compared with that of 
GONG (See in 
the next section); this 
is the result of an 
effective apodization function (limb darkening) which creates a 
narrower spatial response than that of GONG (line-of-sight 
projection).

Figure \ref{year2} has already shown that we could clean the $l=2$ 
LOI data from the $l=0$ mode.  This was done by taking into account 
the 
leakage matrix $\tens{\cal{C}}^{(0,2)}$.  Unfortunately as mentioned 
for the 
$l=5$, the LOI data cannot be cleaned from aliasing degrees because 
the leakage matrice $\tens{\cal{C}}^{(1,6)}$, 
$\tens{\cal{C}}^{(4,7)}$ and 
$\tens{\cal{C}}^{(5,8)}$ are singular.

\vspace{0.5 cm}

\subsubsection{GONG data}
\begin{figure*}[t]
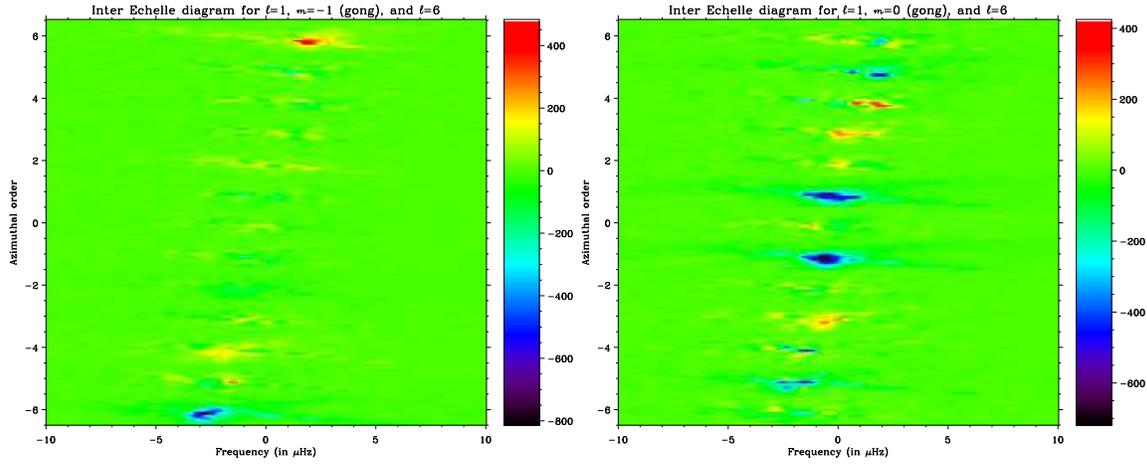

\centerline{\hbox{
\includegraphics[angle=90,width=7.5cm]{echellegong161t.epsi}
\includegraphics[angle=90,width=7.5cm]{echellegong160t.epsi}}
}
\caption{Real part of the inter echelle diagrammes of 1 year of GONG 
data for $l=1$ 
and $l'=6$.  The scale is in m$^{2}\,$s$^{-2}\,$Hz$^{-1}$.  The 
spacing is tuned for $l'=6$.  (Left) For $l=1, m=-1$.  
The strongest correlations are for $m'=\pm 6$ due to $l'=6$.  (Right) 
For $l=1, 
m=0$.  The $l=6$ 
modes are strong in all the panels.  For $m'=\pm 1$ other modes can 
be 
seen as an `asterisk' ridge: the ridge going from bottom right to 
upper left represents the $l=1$ modes; the ridge going from bottom 
left to 
upper right represents the $l=9$ modes; the vertical ridge represents 
the $l=6$ modes. At the far left, the ridge parallel to that of the 
$l=1$ represents the $l=3$ modes.}
\label{gong16} 
\end{figure*}

\begin{figure*}[p]
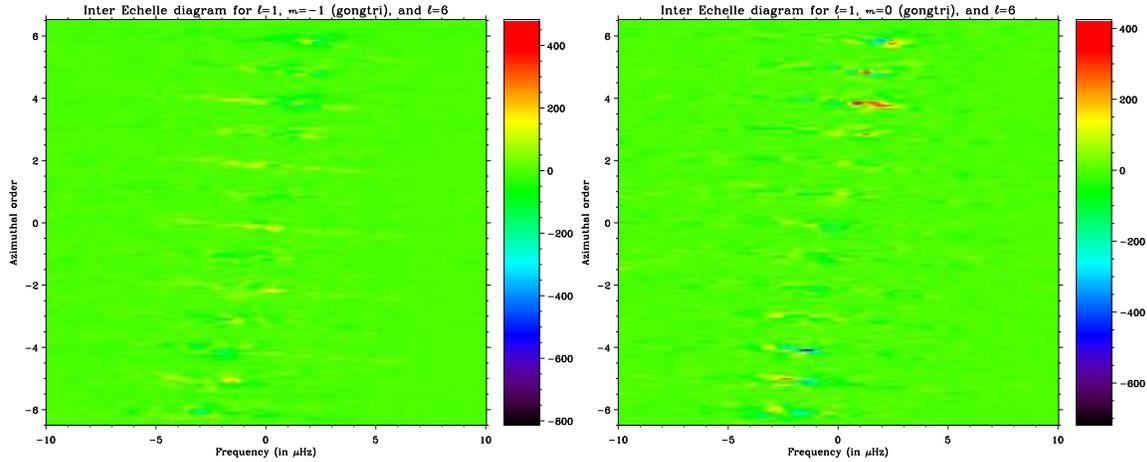

\centerline{\hbox{
\includegraphics[angle=90,width=7.5cm]{echellegongtri161t.epsi}
\includegraphics[angle=90,width=7.5cm]{echellegongtri160t.epsi}
}}
\caption{Real part of the inter echelle diagrammes of 1 year of GONG 
data for $l=1$ 
and $l'=6$ after having applied the inverse of the leakage matrix 
$\tens{\cal{C}}^{(1,6,9)}$.  The scale is in 
m$^{2}\,$s$^{-2}\,$Hz$^{-1}$.
The spacing is tuned for $l'=6$.  (Left) For $l=1, m=-1$.  The faint 
yellow oblique ridges are due to the $l=1$ modes.  (Right) For $l=1, 
m=0$.  The correlations due to the $l=1$ and 6 
modes are almost entirely removed.}
\label{gongtri16} 
\end{figure*}

\begin{figure*}[t]
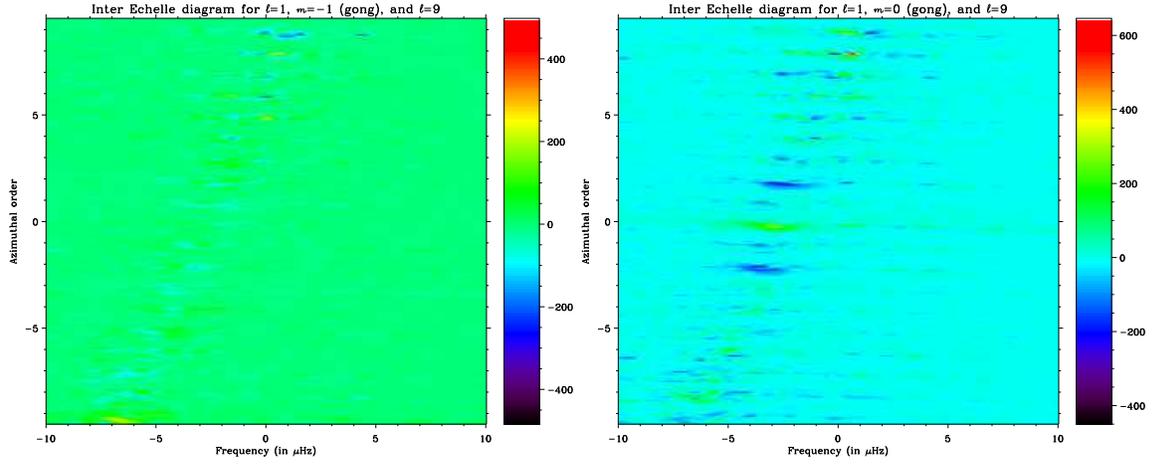

\centerline{\hbox{
\includegraphics[angle=90,width=7.5cm]{echellegong191t.epsi}
\includegraphics[angle=90,width=7.5cm]{echellegong190t.epsi}}
}
\caption{Real part of the inter echelle diagrammes of 1 year of GONG 
data for $l=1$ 
and $l'=9$.   The scale is in cm$^{2}$\,s$^{-2}\,\mu$Hz$^{-1}$.  The 
spacing is tuned for $l=9$.  
(Left) For $l=1, m=-1$.  The strongest correlations are for $m'=\pm 
9$ due to the $l'=9$ modes.  (Right) For $l=1, m=0$.  
The strongest correlations are for $m'=0, \pm 1$ mainly due to the 
$l'=9$ modes.  A faint oblique ridge can be seen for $m'=0$ due to 
the $l=1$ modes.}
\label{gong19} 
\end{figure*}

\begin{figure*}[p]
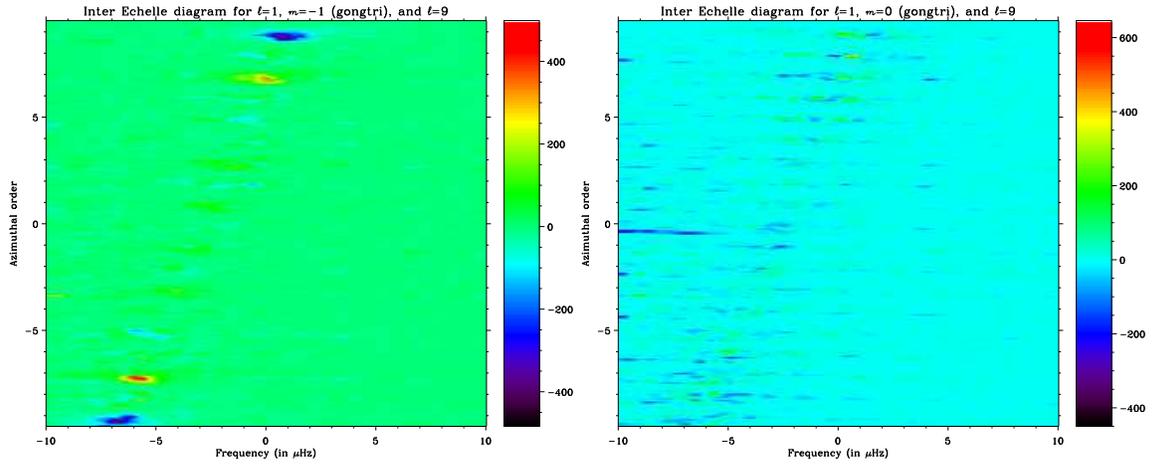

\centerline{\hbox{
\includegraphics[angle=90,width=7.5cm]{echellegongtri191t.epsi}
\includegraphics[angle=90,width=7.5cm]{echellegongtri190t.epsi}}
}
\caption{Real part of the inter echelle diagrammes of 1 year of GONG 
data for $l=1$ 
and $l'=9$ after having applied the inverse of the leakage matrix 
$\tens{\cal{C}}^{(1,6,9)}$.   The scale is in 
cm$^{2}$\,s$^{-2}\,\mu$Hz$^{-1}$.  The spacing is tuned for $l'=9$.  
(Left) For $l=1, m=-1$.  The 
correlations due $l=9$ are not correctly compensated for.  (Right) 
For $l=1, m=0$.  The $l=3$ correlation are also enhanced; they can be 
seen as oblique ridges at the 
left hand side of the diagrammes.}
\label{gongtri19} 
\end{figure*}

Figures \ref{gong16} and \ref{gong19} show the inter echelle 
diagramme for $l=1, l'=6$, and for $l=1, l'=9$.  The correlations are 
clearly visible.  If they
are not taken into account they are likely to bias the frequency 
and splittings of the $l=1$ (See Rabello-Soares and Appourchaux, 
1998, in preparation).  By using the inverse full leakage 
matrix of $l=1,6$ and 9 (i.e. $\tens{\cal{C}}^{(1,6,9)}$, one can 
clean the 
data from these spurious correlations arising both from the aliasing 
degrees and from the $m$ leaks.  Figures \ref{gongtri16} and 
\ref{gongtri19} shows the 'cleaned' inter echelle diagramme which 
should be compared with Figs. \ref{gong16} and \ref{gong19}, 
respectively.  The correlations have been almost entirely removed.  
Some correlations between the $l=1, m=-1$ modes and the $l=6,9$ modes 
are still present.  This is due to the fact that the computation of 
the leakage for $l=1, m=\pm1$ are very sensitive to the subtraction 
of the velocity of the Sun seen as a star.  This is not the case for 
the $l=1, m=0$ modes as they are insensitive to this effect.   
Nevertheless, our imperfect knowledge of the leakage matrix can be 
adjusted in order to remove the residual correlations.  It helped to 
reduce the systematic 
errors of the $l=1$ splitting at high frequencies, mainly where the 
$l=6$ start to cross over the $l=1$ (Rabello-Soares and Appourchaux 
1998, in preparation).  

It is possible to apply this cleaning technique to higher degree 
modes ($l \approx 50-100$.  
In principle, this is feasible, although the plethora of data 
involved 
may be fairly substantial.
For higher degrees, the ridges of the modes in the $(m,\nu)$ echelle 
diagrammes 
are almost 
parallel to
each other.  The idea would be not to use the full Fourier spectra, 
as we did
for the low degree GONG data, but to use only a small frequency range 
of 
about
$\pm$ 30 $\mu$Hz around the target degree.  By doing so, we will not 
only
clean the data from the aliasing degrees but also from the $m$ 
leaks.  In this case, the mode covariance matrix is diagonal.  If we 
assume, wrongly, that the
noise covariance matrix is diagonal, the statistics of each 
cleaned spectrum, for high degree modes, could be approximated by a 
$\chi^{2}$ with 2 degree of freedom.  Neglecting the off-diagonal 
elements of the noise covariance matrix will lead to larger mode 
linewidth, thereby producing underestimated splittings; this 
systematic bias will decrease as the degree.  This approximation will 
produce much less systematic errors than the approximation used by 
Hill \etal (1996) for the original data.  
The use of the suggested approximation for the GONG data may help, at 
the same time, to reduce systematic errors, and to increase computing 
speed for higher degree modes.  The degree where one needs to switch 
between this approximation and the correct analysis needs to be 
determined.

\vspace{0.5cm}

\subsection{Noise covariance matrix measurement}
\subsubsection{LOI/SOHO data}

\begin{figure*}[h]
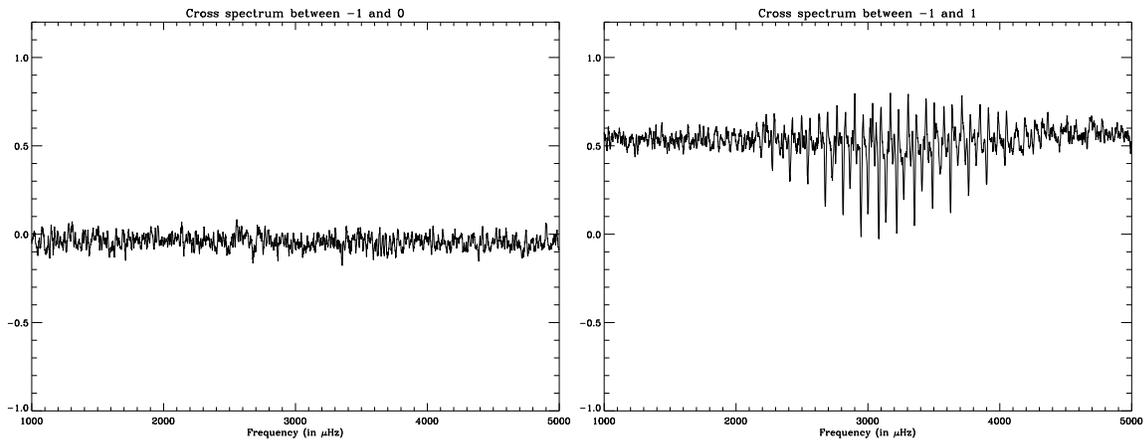

\centerline{\hbox{
\includegraphics[angle=90,width=7.5cm]{noise10-1.epsi}
\includegraphics[angle=90,width=7.5cm]{noise11-1.epsi}}
}
\caption{Real part of the ratio cross spectra for $l=1$. (Left) For 
$m=-1$ and 
$m'$=0.  
There is very little correlation, less than -0.05.  The modes also 
are not seen. (Right) For $m=-1$ and $m=1$.  The noise correlation is 
about 
0.55.  The modes are visible with various positive and negative 
correlation.}
\label{ratio1} 
\end{figure*}

The leakage and ratio matrices have 
similar properties (See Part I).  For example, for a given $l$ there 
is no 
correlation between 
the $m$ for which $m+m'$ is odd for either matrix.  This lack of 
correlation can be seen in Fig. \ref{ratio1} for the ratio cross 
spectra of the LOI data.  The ratio matrix is also useful if one 
wants to reduce the number of noise parameters to
be 
fitted.  This is useful when the noise background, mainly of solar 
origin, varies 
slowly over the p-mode range.  In this case the ratio can be assumed 
to be 
constant over the p-mode range.  For example for $l=1$ we can fit 2 
noise parameters instead of 
3.  When the noise varies in the p-mode range, it is more advisable 
to fit as many noise parameters as required.

\begin{figure*}[!]
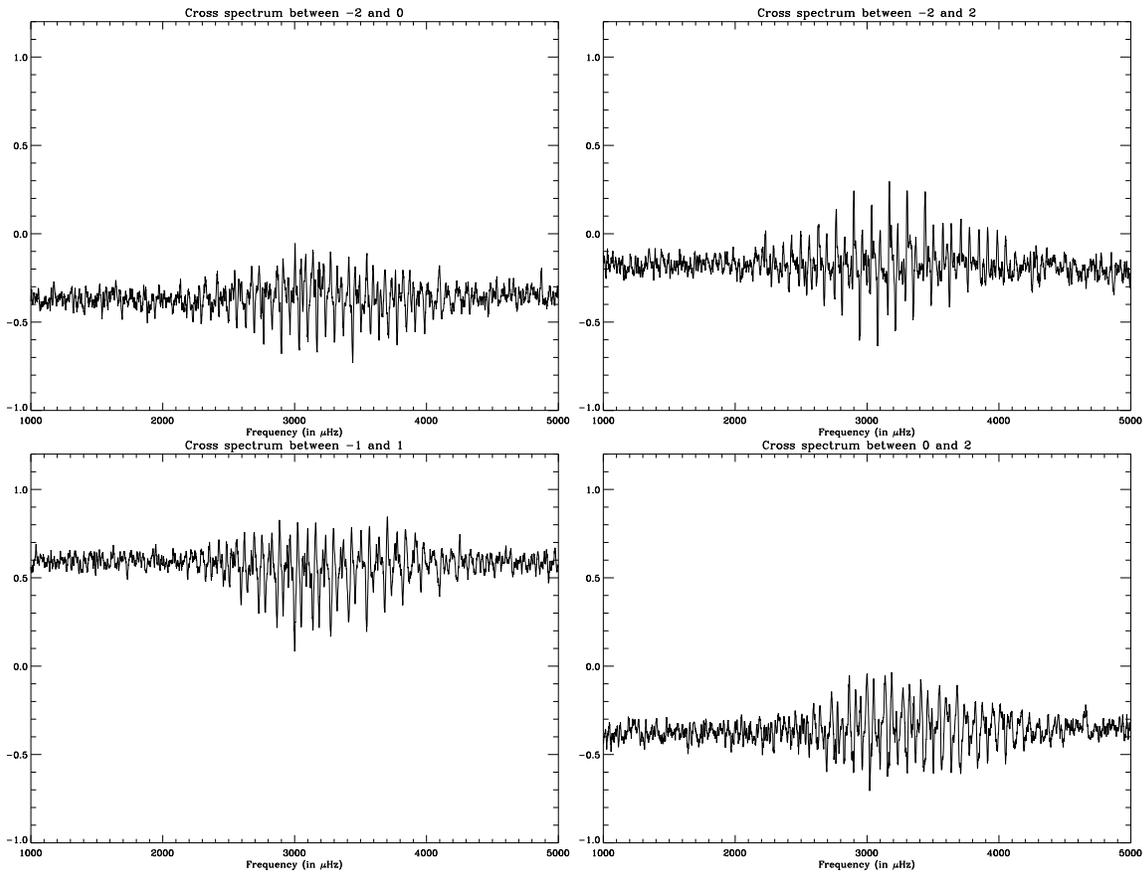

\centerline{\vbox{
\hbox{
\includegraphics[angle=90,width=7.5cm]{noise20-2.epsi}
\includegraphics[angle=90,width=7.5cm]{noise22-2.epsi}}
\hbox{
\includegraphics[angle=90,width=7.5cm]{noise21-1.epsi}
\includegraphics[angle=90,width=7.5cm]{noise220.epsi}}
}}
\caption{Real part of the ratio cross spectra of $l=2$ for the LOI 
data. (Top, left) 
For $m=-2$ and 
$m'=0$.  The noise correlation is about -0.45.
(Top, right) For $m=-2$ and $m=2$.  The noise correlation is about 
-0.20.  (Bottom, left) For $m=-1$ and 
$m'=1$.  The noise correlation is about 0.60.  (Bottom, right) For 
$m=0$ and $m'=2$.  
The noise correlation is about -0.35 same as for $m=-2$ and 
$m'=0$.}
\label{ratio2} 
\end{figure*}

Figure \ref{ratio2} shows the ratio cross spectra of $l=2$ for the 
LOI data.  As this is commonly the case for the LOI, the ratios are 
independent of frequency.  In addition, as outlined in Part I, the 
ratio matrix is very close to the leakage matrix.  Using the values 
of $\alpha$ given for the LOI in Eq. 
(\ref{leakageLOI}) and Fig. \ref{ratio2}, one can see that this is 
the case for the LOI.  The ratio matrix has even the same symmetry 
property as the leakage matrix.  In the case of the LOI, we sometimes 
use the independence of the ratio with frequency to reduce the number 
of free parameters.  It is less straightforward to measure the noise 
correlation for the GONG data than for the LOI data.  We recommend to 
measure it
in between the p modes because that is what the fitting routines will 
determine.

\vspace{0.5 cm}

\subsubsection{GONG data}
\begin{figure*}[!]
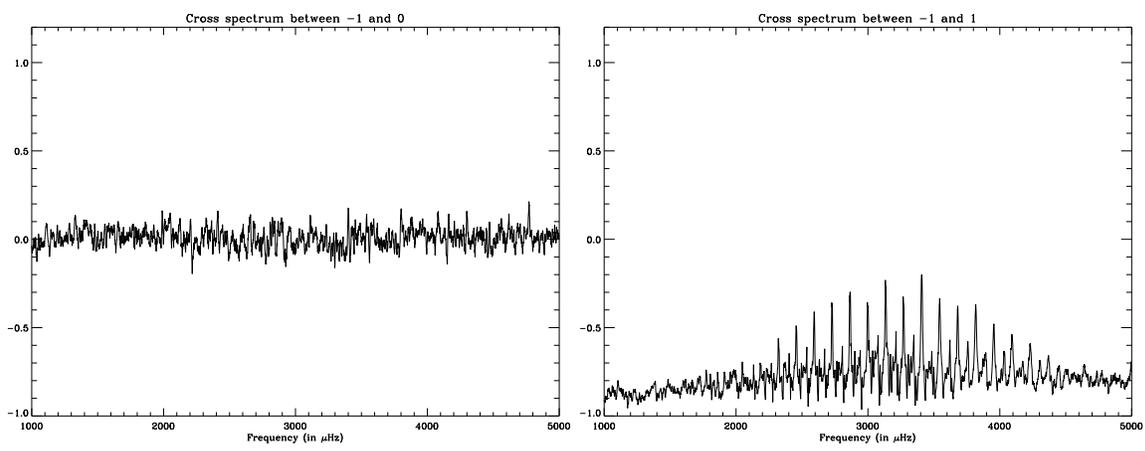

\centerline{\hbox{
\includegraphics[angle=90,width=7.5cm]{gong10-1.epsi}
\includegraphics[angle=90,width=7.5cm]{gong11-1.epsi}}
}
\caption{Real part of the ratio cross spectra of $l=1$ for the GONG 
data. (Left) For 
$m=-1$ and 
$m'$=0.  
As for the LOI, there is very little correlation, less than -0.05.  
The modes also 
are not seen. (Right) For $m=-1$ and $m=1$.  The noise correlation is 
-0.77.  The correlation is negative because of the way the GONG data 
are reduced: the full-disk integrated
velocity has been removed from the signal.}
\label{ratiog1} 
\end{figure*}

\begin{figure*}[!]
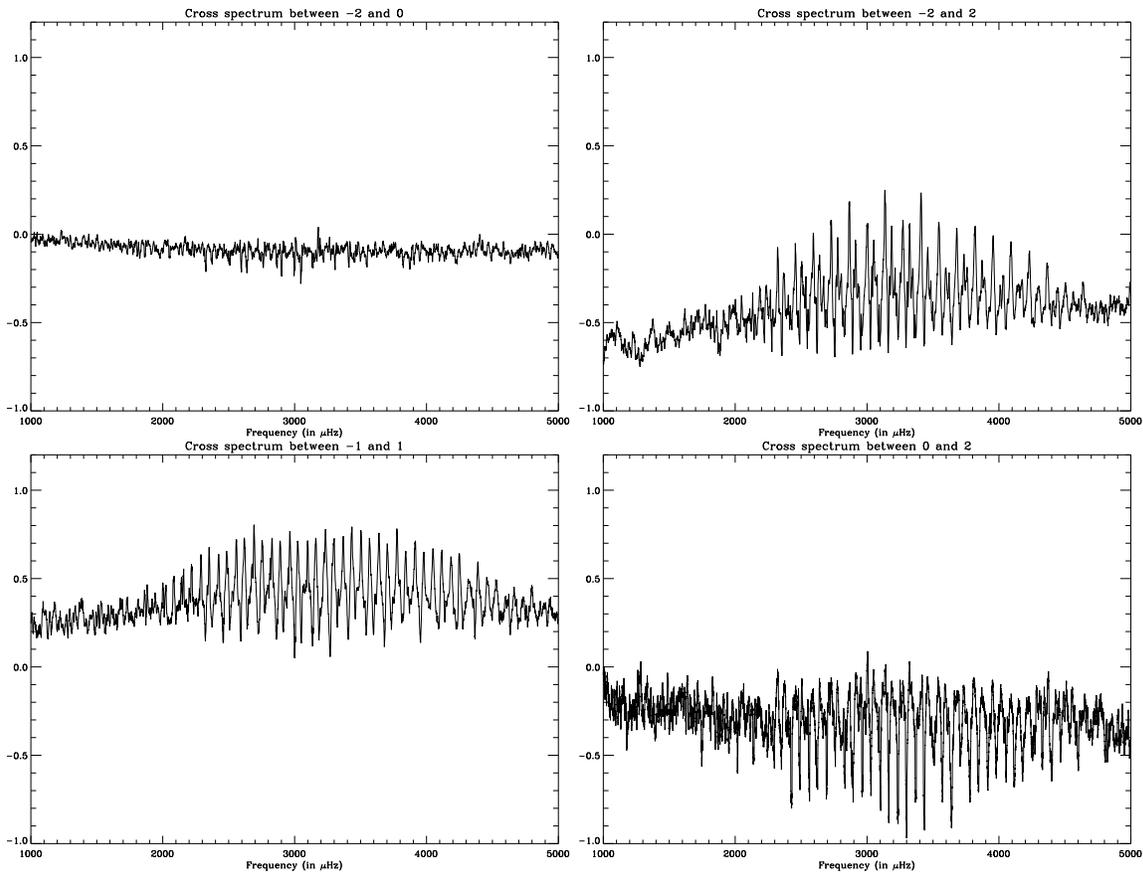

\centerline{\vbox{
\hbox{
\includegraphics[angle=90,width=7.5cm]{gong20-2.epsi}
\includegraphics[angle=90,width=7.5cm]{gong22-2.epsi}}
\hbox{
\includegraphics[angle=90,width=7.5cm]{gong21-1.epsi}
\includegraphics[angle=90,width=7.5cm]{gong220.epsi}}
}}
\caption{Real part of the ratio cross spectra of $l=2$ for the GONG 
data. (Top, left) 
For $m=-2$ and 
$m'=0$.  The noise correlation is rather small and about -0.1.
(Top, right) For $m=-2$ and $m=2$.  The noise correlation is 
frequency dependent, with a typical value at 
3000 $\mu$Hz of about -0.3.  (Bottom, left) For $m=-1$ and 
$m'=1$.  The typical value is 0.4.  (Bottom, right) For 
$m=0$ and $m'=2$.  The typical value is -0.3}
\label{ratiog2} 
\end{figure*}

Figures \ref{ratiog1} and \ref{ratiog2} show the ratio cross spectra 
of $l=2$ for the 
GONG data.  The ratio matrices (as the leakage 
matrices) are not symmetrical.  Apart from the $l=1$, these matrices 
tend to be very close to 
those of the leakage matrices (see previous section).  It is also 
clear that the ratios depend upon frequency, probably due to 
the effect of mesogranulation affecting the spatial properties of the 
noise with frequency.  In this case, we cannot reduce the number of 
parameters to be fitted (see Rabello-Soares and Appourchaux 1998, 
in preparation) as we do for the LOI.

\vspace{0.5 cm}

\section{Summary and Conclusion}
For helping the understanding of fitting p-mode Fourier spectra, we 
have devised 4 new helioseismic diagnostics:
\begin{enumerate}
\item the $(m,\nu)$ echelle diagramme helps visualizing:
\begin{itemize}
\item which degree can be detected,
\item the mode splitting and
\item spurious degrees.
\end{itemize}
\item the cross echelle diagramme helps verifying:
\begin{itemize}
\item the sign of the imaginary parts of the Fourier spectra 
for $m < 0$,
\item the sign of the elements of the leakage matrix of the 
degree $l$,
\item to the first order the theoretical leakage matrix of 
the degree $l$ by applying the inverse of this matrix to the 
data (i.e. cleaning the data from the $m$ leaks) and
\item the cleaning of the data from $m$ leaks.
\end{itemize}
\item the inter echelle diagramme helps verifying:
\begin{itemize}
\item the sign of the elements of the full leakage matrix for 
degrees $l,l'$,
\item to the first order the theoretical leakage matrix of 
the degree $l,l'$ by applying the inverse this matrix to the 
data and
\item the cleaning of the data from aliasing degrees.
\end{itemize}
\item the ratio cross spectrum helps deriving :
\begin{itemize}
\item the amplitude of the noise correlation between the 
Fourier spectra and
\item the frequency dependence of the correlations.
\end{itemize}
\end{enumerate}
These steps will help to evaluate the matrices of the 
leakage, mode 
covariance, ratio and noise covariance directly from 
observations.  The fitting of 
the p modes will be considerably eased by verifying that the 
theoretical knowledge of these matrices is correct.

These steps have been used both on data for which we design the 
spatial filtering (LOI 
instrument), and on data for which we did not design the filtering 
(GONG).  As a matter of fact 
these diagnostics are so powerful that a theoretical knowledge of 
the various matrices is not always necessary to understand the data.  
Very often these diagnostics can also be 
used to find bugs in the data analysis routines (Fig. \ref{error2}).  
Nevertheless, it is advisable to know to the first order the leakage 
and ratio matrices for speeding up the analysis process.  

Last 
but not least, the knowledge of the leakage matrices can be used for 
cleaning the data from $m$ leaks and  from undesired aliasing 
degrees.  This cleaning can 
be performed either when the pixel time series are available (MDI 
data, Toutain \etal, 1998) or more simply 
when only the Fourier spectra are available (LOI, GONG).  
The cleaning has very useful
application for the GONG velocity data for removing 
aliasing degrees from $l=1$, 6 and 9 for instance, and for inferring 
better $l=1$ splittings (Rabello-Soares and Appourchaux 1998, in 
preparation).  The 
cleaning is somewhat easier with Fourier spectra as one does not need 
to reduce large amount of image time series.  It is advisable that, 
in 
the near future, 
this cleaning technique be used for higher degree modes.  This will 
hopefully provide helioseismology with frequency and splitting data 
having much less systematic errors than before.

\begin{acknowledgements} 
SOHO is a mission of international collaboration between ESA and 
NASA.  This work utilizes data obtained by the Global Oscillation 
Network Group (GONG) project managed by the National Solar 
Observatory, a Division of the National Optical Astronomy 
Observatories, which is operated by AURA, Inc. under a cooperative 
agreement with the National Science Foundation.  The GONG data were 
acquired by instruments operated by the Big Bear Solar Observatory, 
High Altitude Observatory, Learmonth Solar Observatory, Udaipur Solar 
Observatory, Instituto de Astrof\'isico de Canarias and Cerro Tololo 
Interamerican Observatory.
Many thanks to Takashi Sekii for constructive comments on the 
manu\-script, and for extensive cyberspace  chatting.  I am grateful 
Mihir Desai for proofreading the English.  Last but not 
least, many thanks to my wife for her patience during the painful 
writing of these 2 articles.
\end{acknowledgements}

\clearpage

\clearpage

\end{document}